\documentclass[11pt,a4paper,iop]{emulateapj}

\usepackage{amsmath,amssymb}
\usepackage{xcolor}
\usepackage[]{graphicx}
\usepackage[english]{babel}

\usepackage[normalem]{ulem}

\usepackage{floatrow}
\usepackage[caption=false]{subfig}

\newcommand{\sigmalos}{\sigma_{\mbox{\tiny $B_{los}$}}}
\newcommand{\sigmassnn}{\sigma_{\mbox{\tiny $\tilde{W}_N$}}}
\newcommand{\sigmassns}{\sigma_{\mbox{\tiny $\tilde{W}_S$}}}

\newcommand{\myvect}[1]{\mathbf#1}
\newcommand{\myhat}[1]{\mathbf#1}
\newcommand{\posi}{\myvect{r}}

\newcommand{\hcpi}[1]{{}{#1}}
\newcommand{\hcpii}[1]{{}{#1}}

\newcommand{\draftgraphicx}{false} 

\shorttitle{Meridional Circulation and the 11-yr Solar Cycle II}
\usepackage{lineno}
\begin{document}

\title{\hcpi{Variational estimation of the large scale time dependent meridional circulation in the Sun: 
proofs of concept with a solar mean field dynamo model}
}

\author{Ching Pui Hung$^{1,2}$, Allan Sacha Brun$^{2}$, Alexandre Fournier$^{1}$, 
        Laur\`ene Jouve$^{2,3}$, Olivier Talagrand$^{4}$, Mustapha Zakari$^{1}$}

\affil{ $^1$ Institut de Physique du Globe de Paris, Sorbonne Paris Cit\'{e}, Universit\'{e} Paris Diderot UMR 7154 CNRS, F-75005 Paris, France\\
 $^2$ Laboratoire AIM Paris-Saclay, CEA/IRFU Universit\'{e} Paris-Diderot CNRS/INSU, 91191 Gif-Sur-Yvette, France\\
 $^3$ Universit\'{e} de Toulouse, UPS-OMP, Institut de Recherche en Astrophysique et Plan\'{e}tologie, 31028 Toulouse Cedex 4, France\\
 $^4$ Laboratoire de m\'{e}t\'{e}orologie dynamique, UMR 8539, Ecole Normale Sup\'{e}rieure, Paris Cedex 05, France}

\begin{abstract}
We present in this work the development of a solar data assimilation method 
based on an axisymmetric mean field dynamo model and magnetic surface data, 
our mid-term goal is to predict the solar quasi cyclic activity. 
Here we focus on the ability of our algorithm to constrain the deep meridional circulation 
of the Sun based on solar magnetic observations. 
To that end, we develop a variational data assimilation technique. 
Within a given assimilation window, the assimilation procedure minimizes 
the differences between data and the forecast from the model, 
by finding an optimal meridional circulation in the convection zone, 
and an optimal initial magnetic field, via a quasi-Newton algorithm. 
We demonstrate the capability of the technique to estimate the meridional flow 
by a closed-loop experiment involving 40 years of synthetic, solar-like data. 
By assimilating the synthetic magnetic proxies annually, we are able to reconstruct a (stochastic) 
time-varying meridional circulation which is also slightly equatorially asymmetric. 
We show that the method is robust in estimating a flow whose level of fluctuation can reach 
30\% about the average, and that the horizon of \hcpii{predictive capability} of the method is of the order of 1 cycle length.

\end{abstract}

\keywords{Sun: meridional circulation, activity, dynamo, methods: numerical, data assimilation}

\section{Introduction} 
\label{sec:sec_intro}
\subsection{Solar activity: Observations and Models}
\label{subsec:SOM}
The Sun is an active star. 
Solar activity includes surface magnetic variability, solar eruption, 
coronal activity and its effects on planets through magnetic disturbances. 
The Sun is a nonlinear system and it is 
a real challenge to predict its future activity.
Since solar activity impacts space-weather, 
which in turn alters our modern technology-based society significantly, 
it has become increasingly important to obtain good solar predictions. 
The most common index to quantify solar activity is the sunspot number (SSN). 
\citep[For recent discussion of SSN, see][]{CletteLefevre2012JSWSC, Clette2014SSR, 
Svalgaard2016SolarPhys,Vaquero2016SolarPhys}.  
Sunspots are dark areas on the solar disc, 
where mostly vertical magnetic field of $\sim 3$kG peak values, is present \citep{stix2002book}. 
In 1850, Rudolf Wolf introduced the 
relative sunspot number $R_z=k(10g+s)$, where $g$ is the number of sunspot groups, 
$s$ is the number of individual sunspots, $k$ is a constant to account for the differences in observations from 
various observers and astronomers \citep{Wolf1850MiZur}. 
The corresponding sunspot series started in 1749. 
In addition to the sunspot number, the surface magnetic field of the Sun is also an important observable.
Observations of solar magnetic field can at least be traced back as early as in 1908 
through the pioneering observations of \cite{Hale1908ApJ}. 
Systematic, daily observations of solar magnetic field 
over the solar disk started in early 1970s at the Kitt Peak National Observatory, 
with synoptic maps nearly continuously measured from early 1975 through mid 2003 \citep{Hathaway10}. 
Tracing the surface radial magnetic field as a function of time and latitude, 
averaged over longitude, enables to construct the so-called butterfly diagram. 
It shows the position where sunspots appear during a solar cycle, 
and exhibits their phase relationship with the strength of the polar field. 
One of the most prominent features of the solar activity is the quasiperiodicity of the sunspot cycles of 11 years. 
Those cycles, however, vary in both their period and their amplitude \citep[for more recent time series, 
\hcpii{consult}][]{Svalgaard2015Solarphy}. 

The long-term \hcpii{(multi-decadal)} variation shows randomness, 
but with highs in sunspot number every 7 or 8 cycles 
\citep{GleissbergObs1939,Usoskin13}. 
Furthermore, sometimes the solar activity is broken up; 
the periods of such depression are called grand minima.
The significant modulation of solar activity raises questions regarding its predictability. 
Studies suggest that the predictability also depends on whether the source of 
the variability of solar dynamo is deterministic or not 
\citep{Ossendrijver02, Tobias1998ApJ, Brandenburg2008AN}; 
even a weak stochastic perturbation can lead to a loss in predictability \citep{Bushby2007ApJ}. \par 

\hcpii{Dynamo models based on magnetohydrodynamics (MHD) are a common class of models established 
to account for the solar activity \citep{Charbonneau2010lrsp}. }
The model used in our assimilation framework \hcpii{(to be discussed below)} 
is a dynamo model based on the mean field induction equation, 
in spherical coordinates with azimuthal symmetry. 
\hcpii{Its} mechanism was proposed by \cite{Babcock61} and elaborated by \cite{Leighton69}. 
This model can also account for Joy's law \citep{Haleeta1919ApJ}. 
Numerical studies of the \hcpii{so-called} Babcock-Leighton dynamo model are widely established 
\citep[e.g.][and references therein]{Dikpati99,JouveMC07}.  
\par
In this flux transport solar dynamo model, 
the meridional circulation in the convection zone is the key ingredient determining 
the length of the solar cycle. 
The effects of the meridional circulation
on the magnetic cycle and magnetic field are investigated in detail in 
\cite{JouveMC07}, \cite{hazra2014ApJ} \hcpii{and \cite{Belucz2015ApJ}}. 
\hcpii{A meridional circulation with one cell per hemisphere is frequently used as a reference
in this model to account for the cycle length, maxima and phase relationship in solar activity.
Additional cells in radius and latitude can result in different effects on the advection of magnetic field and cycle length. 
A 2-cell in radius  meridional flow implies 
the presence of a return flow at mid-depth which slows down the transport of the flux from the surface to the tachocline,
resulting in a longer cycle length. The flow becomes poleward at the tachocline thus introducing a poleward branch in 
the time-latitude plot of the toroidal flow at the base of the convection zone. 
The toroidal field at the base is weaker 
than that of the unicellular case as the polar fields are advected from the bottom at low latitudes rather than being brought 
from the poles. 
On the other hand, for a dynamo model with 
a 2-cell in latitude (in each hemisphere) meridional flow, the cycle length is shorter than in the unicellular case because of the 
shorter primary conveyor belt, while a poleward branch in the toroidal field at the tachocline is also present, 
as in the 2-cell in radius case. 
For a larger number of latitudinal cells, 
the toroidal field at the base is also weaker than in the unicellular case as the dynamo is confined to 
low latitudes where the differential rotation is smaller \citep{Belucz2015ApJ}.
It is also found that the influence of having several radial cells on the model is stronger than 
that of adding cells in latitude \citep[see][for details]{JouveMC07}. 
} \par
\hcpii{While a flux transport dynamo model with unicellular meridional circulation 
is commonly used to account for the 11-yr solar activity, 
recent estimate of meridional circulation from helioseismology below the solar surface suggests the possibility of 
more complex flow structures. 
For example, in \cite{Zhao13}, a meridional circulation with 2 cells in the radial direction is reported, 
though the errors of the estimate below 0.80$R_{\odot}$ ($R_{\odot}$ is the solar radius) 
are considerably higher than that at the surface. 
In \cite{Schad13}, more complicated structure like 2 cells in radius and 4 cells in latitude is suggested, 
based on perturbation of Solar p-modes eigenfunctions by meridional flow. 
Submerged meridional cell has been discovered by local helioseismic technique of ring diagram analysis of MDI data 
from 1998-2001, which disrupts the orderly poleward flow and equatorial symmetry in those years \citep{Haber2002ApJ}.
Time distance helioseismic measurements using GONG data also suggest multicellular large scale meridional flow in 
the convection zone \citep{Kholikov2014ApJ}. 
In summary, there is no unique conclusion on the meridional flow structure in depth, which also raises the interest 
of estimating the meridional flow with an independent method resting on a dynamo model.}

\subsection{Solar Prediction and Data Assimilation methods}
\label{subsec:SPDA}
\hcpii{Because of the irregular nature of the Solar activity discussed above, }
a wide range of solar prediction methods are developed, 
from the studies of geomagnetic precursors 
to extrapolation methods based on time series analysis of 
the past activity and correlation studies \citep{Hathaway1999JGR}, 
and to more \hcpii{sophisticated} methods using numerical models which simulate the evolution of the system 
on the basis of the relevant physical equations. Such numerical models require the definition of 
\hcpii{adequate} initial conditions 
which are obtained through the technique of data assimilation 
\citep{Petrovay10, Dikpati06, pesnell2016SW}. 
Data assimilation is an emerging technique in solar cycle and activity prediction, 
which is a way to incorporate observations in numerical models \citep{Brun2007AN}. 
Suppose some solar observations are available on a time interval. 
By controlling the initial condition and key control parameters of a numerical model, 
\hcpii{the task of a data assimilation method} is to obtain a model trajectory 
which can best account for the observations. 

Modern data assimilation techniques can be split into 
two general classes, sequential and variational. 
The Kalman filter and Ensemble Kalman Filter (EnKF) 
are common methods for the sequential class, 
and make use of observations on the fly, as soon as they are available. 
For the variational approach, by controlling selected parameters of 
the physical model, an optimal fit of data is obtained over the entire time window, 
making use of all the observations available. 
A common example is 4D-Var, 
in which the minimization of the objective function can be implemented 
by the development of an adjoint model \citep{Fournier10, Talagrand2010variational}. 
\hcpii{The respective
 merits and drawbacks of the sequential and variational approaches have been discussed at length 
 (see e.g. \cite{Fournier10}, \S 2.2.3 and references therein). Suffice it to say here 
that both lead to similar answers (identical in the linear case with Gaussian error statistics) 
and that a sequential method is generally easier to implement than a variational 
 method (which requires the implementation of the adjoint model). 
 The variational approach is more flexible, 
and it uses all the observations available over a given 
 time window to define an optimal initial set-up at the beginning of the window.} 
For sequential data assimilation, use of EnKF assimilation in analysis or prediction of solar activity, 
for example, is illustrated by \cite{Kitiashvili08} and \cite{Dikpati14}.
On the other hand, the use of variational data assimilation method 
with solar dynamo models is illustrated by \cite{JouveAssimi11}.
In that paper, an $\alpha \Omega$ mean field dynamo model defined 
on a Cartesian coordinates system is adopted. 
The corresponding adjoint model is developed, followed by a twin experiment 
which successfully estimates the spatial dependence of the physical ingredients of the model,
such as the profile and strength of the $\alpha$-effect. 
Similar developments based on a flux transport dynamo model in axisymmetric
spherical coordinates are presented in \cite{Hungetal2015ApJ} (\hcpi{hereafter} Paper I) 
to estimate the steady meridional flow of the model with synthetic magnetic observations,
as a first step towards predicting the solar cycle. \\
In this study, we are going to extend the framework developed in Paper I, 
by adding the initial conditions to the control vector and estimating a time dependent meridional circulation.
\par

In \hcpi{Paper I},
we included 
the meridional circulation as the main control parameter of our data assimilation pipeline. 
We verified that the variational assimilation method is capable of estimating 
the meridional circulation of the model by minimizing the misfit 
between synthetic magnetic observations and model trajectory. \hcpii{Again, 
  this study assumed a steady meridional circulation.}
In reality, the solar cycle is significantly modulated, and the meridional flow is fluctuating 
\citep{Ulrich2010, BasuMC23cycle10, Komm2015SolarPhys}, 
\hcpi{\citep[for more about observations of meridional flow, see ][]{Haber02, Haber2003ESASP, 
Zhao04, Zhao2012ApJ, Zhao13, Svanda2007ApJ, Svanda2008ApJ, Schad13, Upton14}}, 
so the next step of development is to verify the applicability of the method to capture the variability of 
the modulated activity. 

Similar studies were performed recently, for example, by \cite{Dikpati14} \hcpi{and \cite{Dikpati2016ApJ}.
In \cite{Dikpati14},} a numerical experiment was used to reconstruct the time-varying amplitude of the flow, 
by applying the EnKF to the Babcock-Leighton flux transport dynamo model. 
In this work,  we apply a variational data assimilation method, 
to reconstruct a time varying meridional circulation, 
by ingesting synthetic observations produced by a dynamo model 
with a meridional flow modulated both in amplitude and shape. 
\par
We present our work as follows. 
In Sec. \ref{sec:method}, we describe the motivation and methodology of the assimilation 
framework.
In Sec. \ref{sec:results}, we present the results of the numerical experiment.
We discuss the results of hindcasting in Sec. \ref{subsec:Analysis}. 
Then we investigate the \hcpii{predictive capability} of the assimilation procedure 
and of the model in Sec. \ref{subsec:prediction}.
Furthermore, we test the robustness of the procedure by inverting the synthetic observations 
based on a meridional flow with different levels of fluctuations (Sec. \ref{subsec:fluctlv}). 
We summarize and discuss our results in Sec. \ref{sec:summary}. 
\hcpi{Along with Paper I,} in the Appendix, we describe the Babcock-Leighton 
mean field dynamo model (Sec. \ref{subsec:BLmodel}), 
and we include some details about the algorithm 
which incorporates the initial condition 
in the assimilation procedure (Sec. \ref{sec:asscov}). We finally give a brief analysis of 
the observation of the flow at the surface of the Sun (Sec. \ref{sec:anassn}). 

\section{Methodology}
\label{sec:method}

\subsection{Generation of synthetic data based on a dynamo model with a time varying meridional circulation}
\label{subsec:modeleq}

We presented in \hcpi{Paper I} 
a first step toward predicting 
future solar activity levels using \hcpii{variational} data assimilation.
As a proof of concept, we performed twin experiments for which the assimilated
 data were produced by the flux-transport (Babcock-Leighton) model itself. 
Details on the model and its numerical implementation can be found in Appendix \ref{subsec:BLmodel}. 
\hcpii{The system is axisymmetric and we express the magnetic field $\mathbf{B}(\mathbf{r},t)$ 
as sum of its toroidal and poloidal component, 
and the latter is further expressed as the curl of a vector potential with the axisymmetric assumption: }
\begin{equation}
\label{eq:axisymB}
           \mathbf{B}(\mathbf{r},t)= B_{\phi}(\mathbf{r},t)\mathbf{e_{\phi}}+\nabla \times [A_{\phi}(\mathbf{r},t)\mathbf{e_{\phi}}],
\end{equation}
\hcpii{where $\mathbf{e_{\phi}}$ is the azimuthal unit vector, 
the first and second term are the toroidal and poloidal component of the magnetic field respectively, 
and $A_{\phi}\mathbf{e_{\phi}}$ is the vector potential of the poloidal field. 
The model equations are partial differential 
equations describing the time evolution of $A_{\phi}$ and $B_{\phi}$.}
\hcpii{This model is very similar to the one in Paper I, except that (i) the meridional flow 
(defined with $\psi$) is steady in Paper I but time dependent in this work, 
and (ii) the diffusion profile is slightly modified here compared with that in Paper I.}
The axisymmetric meridional 
 circulation is described using a stream function $\psi(\posi,t)$, in which $\posi$ and
 $t$ denote position in the meridional plane and time, respectively. 

Since the flux-transport
 model we adopted was based on a constant meridional circulation, the 
 regular and periodic synthetic activity it generated 
 lacked some of the salient features
   of solar activity, namely its variability in cycle length and amplitude.  
\hcpii{In fact, the duration of the 23 sunspot cycles since 1749 distribute broadly about $11\pm3$ years.}

To be able to account for these important observational facts, we make the meridional 
 flow of our flux-transport model time-dependent, and write the corresponding
 stream-function $\psi(\posi, t)$ as the sum of a constant (background) 
term $\overline{\psi}(\posi)$ and a time-dependent term (of zero mean) $\psi'(\posi,t)$

\begin{equation}
\psi(\posi, t) = \overline{\psi}(\posi)+ \psi'(\posi,t). 
\end{equation}

\begin{figure}[!ht]
\includegraphics[draft=\draftgraphicx,width=0.75\columnwidth]{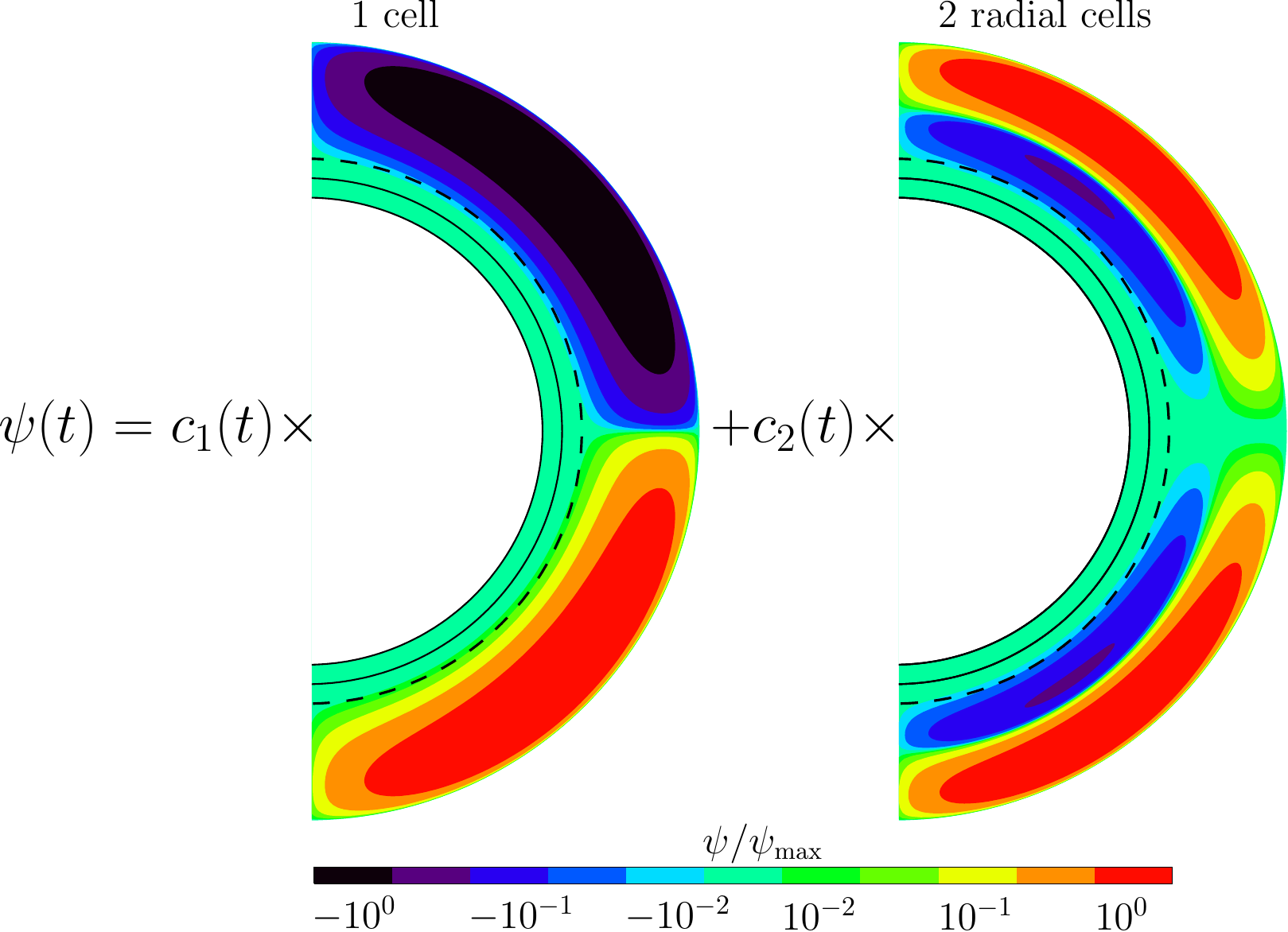}
\caption{Stream functions of those two components of the meridional circulation used 
to generate synthetic observations. The left one ($\psi_1$) is the stream function for
unicellular flow, the right one ($\psi_2$) is the stream function for the equatorially antisymmetric flow. 
Note that the equatorial parity of the stream function is opposite to that of the corresponding flow.
} 
\label{fig:streamsymanti}
\end{figure}

In this study, $\overline{\psi}$ corresponds to an equatorially anti-symmetric,  
one-cell per hemisphere, constant flow whose maximum surface  
amplitude is $v_0=22.3~$m s$^{-1}$. This flow pattern will be 
denoted $\psi_1$ henceforth, 
and its streamlines are shown in the left panel of Fig.~\ref{fig:streamsymanti}. The integration of the model with 
 $\overline{\psi}$ alone leads to a regular activity of period $11.5$ years. 

The fluctuating part
 $\psi'(t)$ comprises two components, whose amplitude is time-dependent:  
 the first is $\psi_1$ and the second ($\psi_2$ henceforth) corresponds to an equatorially symmetric, 
two cells per hemisphere (on the meridional plane, one radial node) flow, shown in the right panel 
 of Fig.~\ref{fig:streamsymanti}. As indicated in this figure, the 
 total flow is therefore a combination of two components and can be written
 as 
\begin{equation}
\psi(\posi, t) =  c_1(t) \times \psi_1 (\posi) + c_2 (t) \times \psi_2 (\posi). 
\label{eq:modelflow}
\end{equation}
We specify the explicit expression of $\psi(\posi, t)$ in this case in terms of its expansion on 
a chosen set of basis functions in Appendix \ref{subsec:BLmodel}.
The coefficients $c_1$ and $c_2$ are constructed as follows 
\begin{eqnarray}
c_1(t)  &=& 1 + A_1 F[\delta_1(t)], \label{eq:c1}\\
c_2(t)  &=& A_2F[\delta_2(t)], \label{eq:c2}
\end{eqnarray}
in which each $\delta_i(t)$ is a random number (drawn from a uniform distribution) 
 whose amplitude
 is normalized so that $\delta_i = 1$ implies a maximum surface velocity 
 equal to $v_0$. 

\begin{figure}[!ht]
\includegraphics[draft=\draftgraphicx,width=\columnwidth]{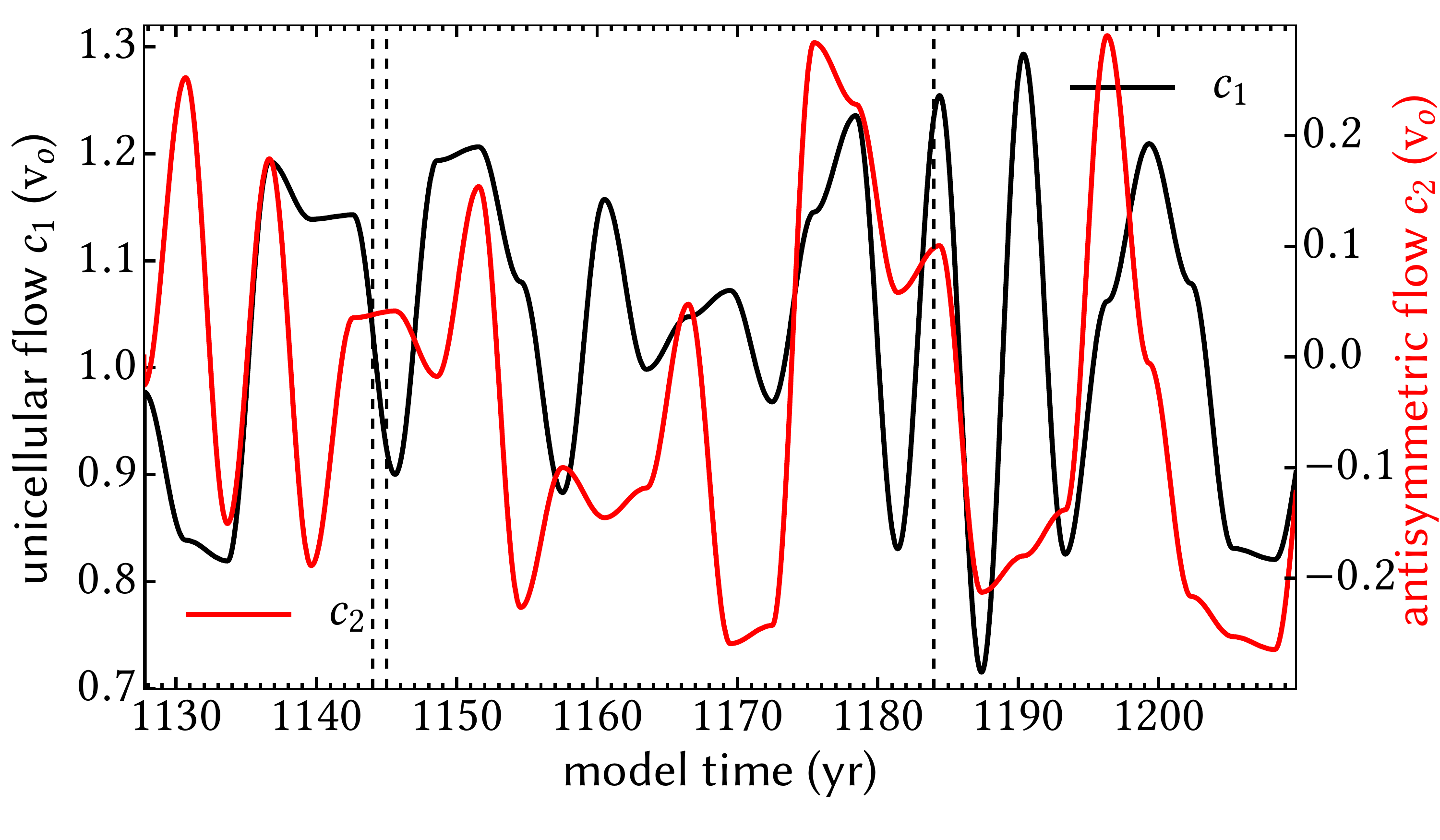}
\caption{Stream function coefficients $c_1$ and $c_2$ adopted in the dynamo model 
to generate the synthetic data, as a function of time. 
The coefficients are normalized such that the corresponding maximum surface flow is  $v_o$. 
The two vertical broken lines on the left mark a typical 
one year sampling window of data assimilation, and the leftmost and rightmost broken lines indicate the whole course of 40 year assimilation. } 
\label{fig:timevaryflow}
\end{figure}

\begin{figure}[!ht]
\includegraphics[draft=\draftgraphicx,width=\columnwidth]{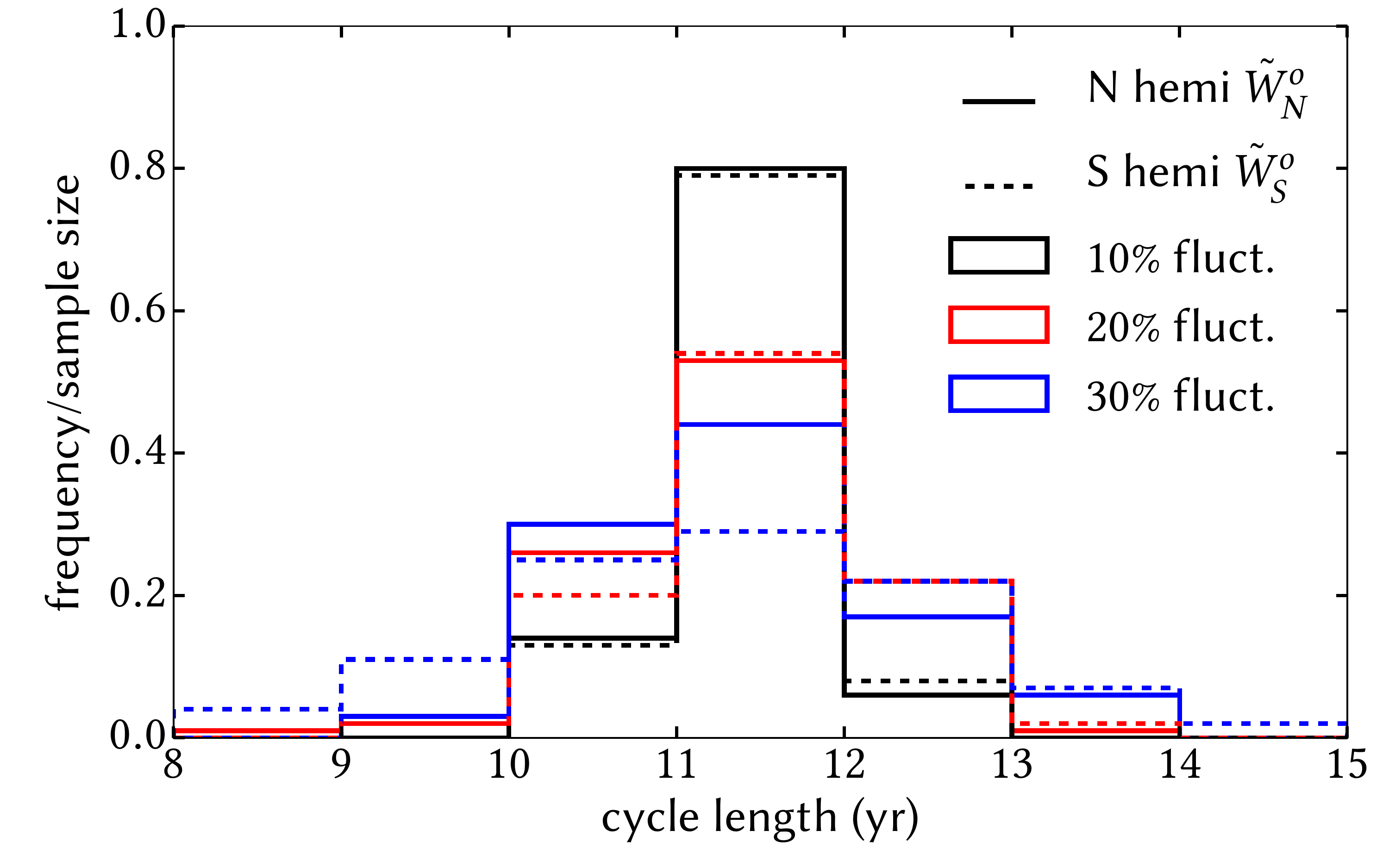}
\caption{Histograms of cycle length of the dynamo model over 100 (synthetic) sunspot cycles, 
with $10\%$ (black), $20\%$ (red) and $30\%$ (blue) 
fluctuation in the meridional flow. 
Solid (resp. dashed) lines refer to statistics in the Northern (resp. Southern) hemisphere. }
\label{fig:ssncycle}
\end{figure}
Substituting Equations \eqref{eq:c1} and \eqref{eq:c2} into \eqref{eq:modelflow}, 
we see that in this study, the time independent part 
$\overline{\psi}(\posi)$ is $\psi_1 (\posi)$, 
and the time dependent part is 
\hcpi{
\begin{equation}
\label{tdsf}
 \psi'(\posi,t)=A_1 F\left[\delta_1(t)\right]\psi_1 (\posi) + A_2F\left[\delta_2(t)\right]\psi_2 (\posi).
\end{equation}
}
 \hcpi{The width $\tau_i$ of the interval between two consecutive
 values of  $\delta_i$ 
 is chosen based  on the available observational evidence. 
 As shown in Appendix~\ref{sec:anassn}, a spectral decomposition of the solar surface flow
  inferred by \cite{Ulrich2010} shows that the equatorially symmetric flows
are dominant with respect to their antisymmetric counterparts. 
The auto-correlation times of the amplitudes vary 
 from $\sim 1$ year for the antisymmetric modes
to $3$ years and more for the symmetric modes. 
 In this study, for the 
 sake of simplicity,  we shall take that time to be $3$ years for both families, and consequently set $\tau_1=\tau_2=3$~years. 
}

\begin{figure}
\centering
{\includegraphics[draft=\draftgraphicx,width=\columnwidth]{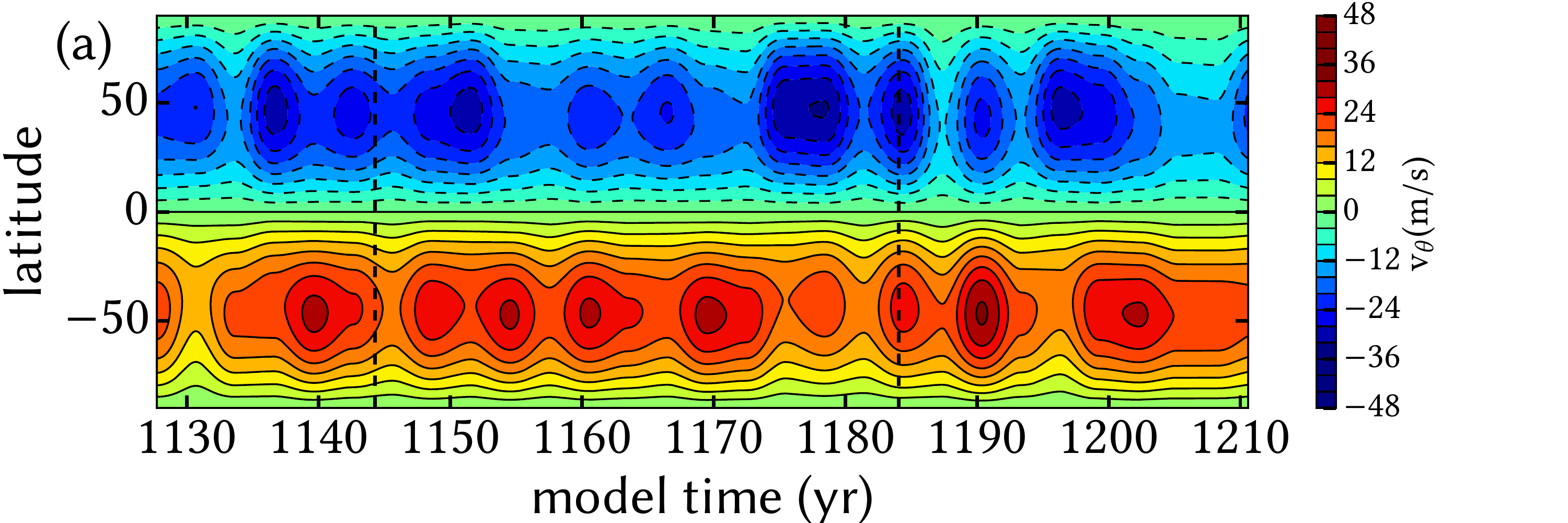}}\\
{\includegraphics[draft=\draftgraphicx,width=\columnwidth]{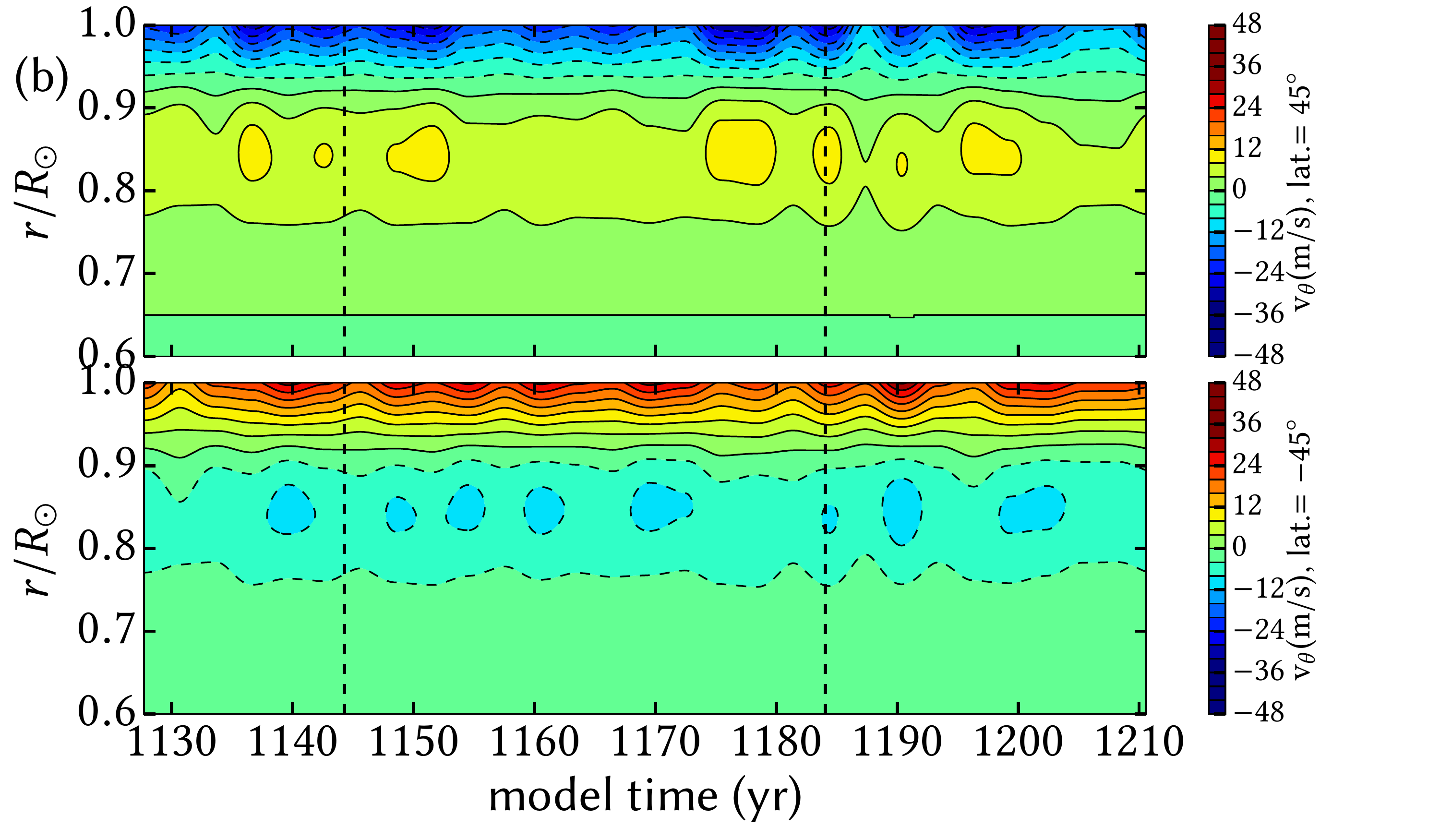}}
\caption{(a) Latitudinal component of the flow at the surface 
as a function of time, in the case of a fluctuation of the meridional flow 
characterized by $A_1=A_2=0.3$. (See text for details.)
The assimilation period in the numerical experiment which follows 
is indicated by the dashed vertical lines. 
The sign convention is positive for a flow due south. 
(b) Latitudinal component of the flow in time-radius contour plots at 
latitude $45^{\circ}$ (top) and $-45^{\circ}$ (bottom); 
same flow setup as in (a).}
\label{fig:uthetacontour}
\end{figure}

\hcpi{
 We next interpolate in time between two consecutive values 
  of $\delta_i$ using a sine function and this interpolation is symbolized  
 by the $F$ operator in Equations~\eqref{eq:c1}, \eqref{eq:c2}.
}
\hcpi{To explicitly define $F$, for any nonnegative integer $n$, suppose random numbers $\delta_{i,n}, i=1,2$, 
are generated at $t=n\tau_i$, then 
\begin{equation}
\begin{split}
F\left[\delta_i(t)\right]=\frac{1}{2}\left\{ \delta_{i,n} + \delta_{i,n+1}  
    + \left(\delta_{i,n}- \delta_{i,n+1}\right)\cos\left[\pi (t/\tau_i-n)\right]   \right\}, \\ 
  \text{for $n\tau_i \leq t < (n+1)\tau_i$}. 
\label{eq:defF}
\end{split}
\end{equation}
}
\hcpi{ 
Figure~\ref{fig:timevaryflow} shows an 
example of realization of $(c_1,c_2)$, for which the chosen level of fluctuation 
amounts to $30$\% of the mean flow (in other words, $A_1=A_2=0.3$ and 
the maximum surface velocity that can originate from $\psi_2$ alone is $7$~ms$^{-1}$, 
and the maximum total fluctuation at the surface 
(from $\psi'(\posi,t)$) can reach $\sim 14$~ms$^{-1}$  ).} \par
\hcpi{ 
The level of fluctuation in $\psi'$ controls the amount of variability in the simulation, 
which can be assessed statistically. }\par 
\hcpi{ 
We show the histograms of cycle duration of the model 
for different fluctuation levels $A_1$ ($A_2$) in Figure \ref{fig:ssncycle}, namely $10$, $20$, and $30$ \%. 
\hcpi{The cycle length is defined by the time between two consecutive minima of the modeled magnetic proxy 
which will be defined shortly after [Equation \eqref{eq:bssn} and \eqref{eq:bssnsn}].}
Each corresponding model has been integrated for a long enough time to enable $100$ cycles to be achieved. 
The statistics shown here are separated into their Northern and Southern contributions. 
For a perturbation of flow speed of  30\%, the cycle length varies from $\sim 9$ to $\sim 14$ years, 
which is in reasonable agreement with observations based on the available records of the $23$ cycles at our disposal. 
 
Unless otherwise stated, we will use this fluctuation level of 30\% in the remainder of this study. 
An example of realization of the meridional flow 
is shown in Fig. \ref{fig:uthetacontour} for the $\theta$ component of the flow,
given the $c_1$ and $c_2$ already displayed in Fig.~\ref{fig:timevaryflow}. 
\hcpii{The meridional flow is dominated by unicellular structure in each hemisphere, 
with equatorial asymmetric fluctuations. This meridional circulation is chosen for our numerical tests in this work, 
as unicellular structure is observed mostly \citep{Ulrich2010, BasuMC23cycle10}, 
though helioseismological studies suggest the presence 
of counter cells in the convection zone \citep{Haber02,Zhao2013ApJ29Z, Schad13}.}
In particular, we present the surface flow in Fig. \ref{fig:uthetacontour} (a), 
and note \hcpii{again that similar} time variability is also \hcpii{reported} in the Sun 
\citep[eg.,][]{Ulrich2005, Komm2015SolarPhys}.}

\begin{figure}[!ht]
\centering
\includegraphics[draft=\draftgraphicx,width=\columnwidth]{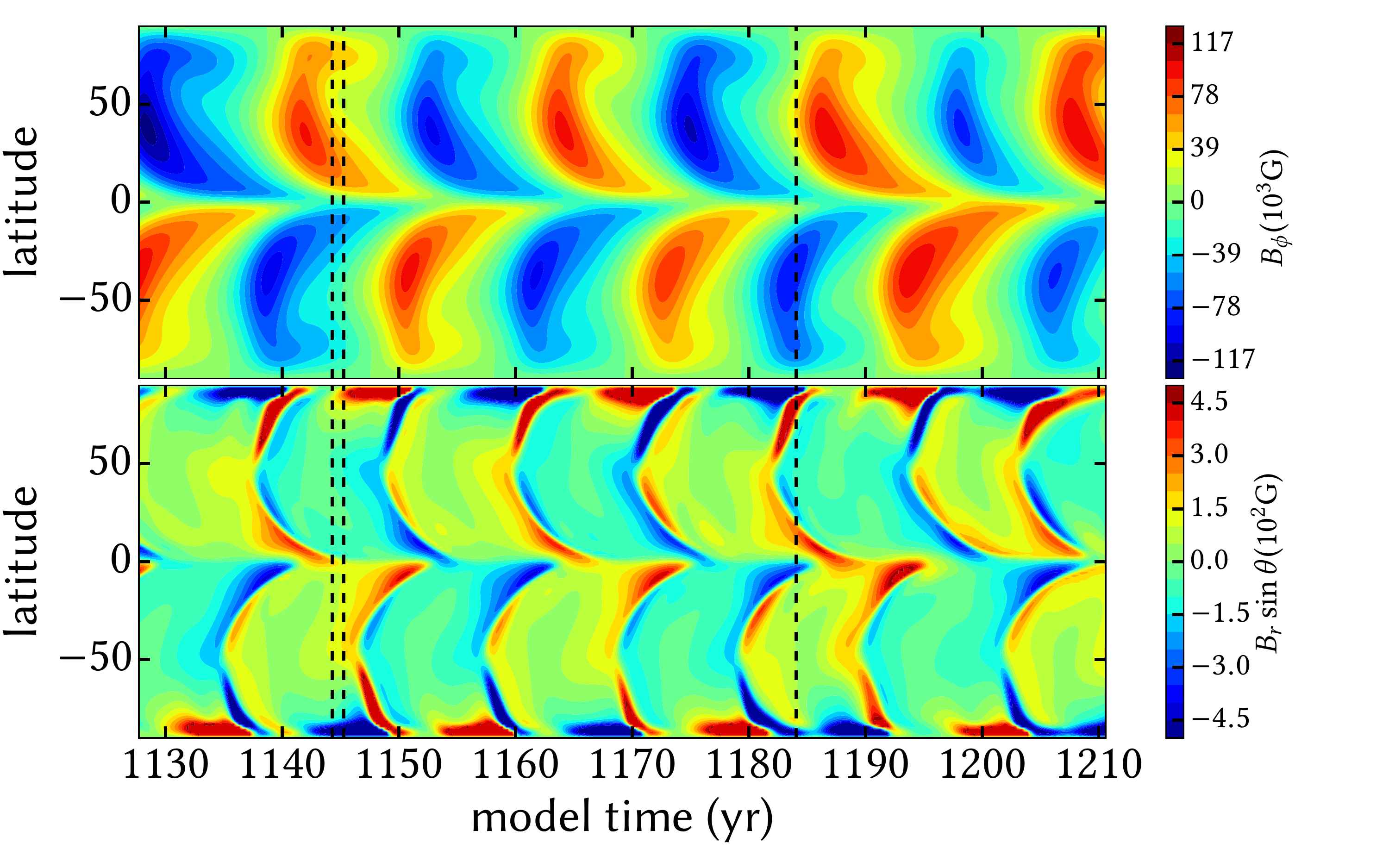}
\caption{Top: time-latitude representation of the toroidal field at the tachocline. 
Bottom: time-latitude evolution of the magnetic field in the line of sight at the surface. 
}
\label{fig:syntheticfield}
\end{figure}

The plots show the asymmetry of the flow about the equator. 
The corresponding simulated magnetic field is shown in Fig.~\ref{fig:syntheticfield}, 
which shows the advection of the toroidal field toward the equator at the base of the convection zone, 
and the polar branch at the surface shows the radial field is advected polewards.

\begin{figure}[!ht]
\includegraphics[draft=\draftgraphicx,width=\columnwidth]{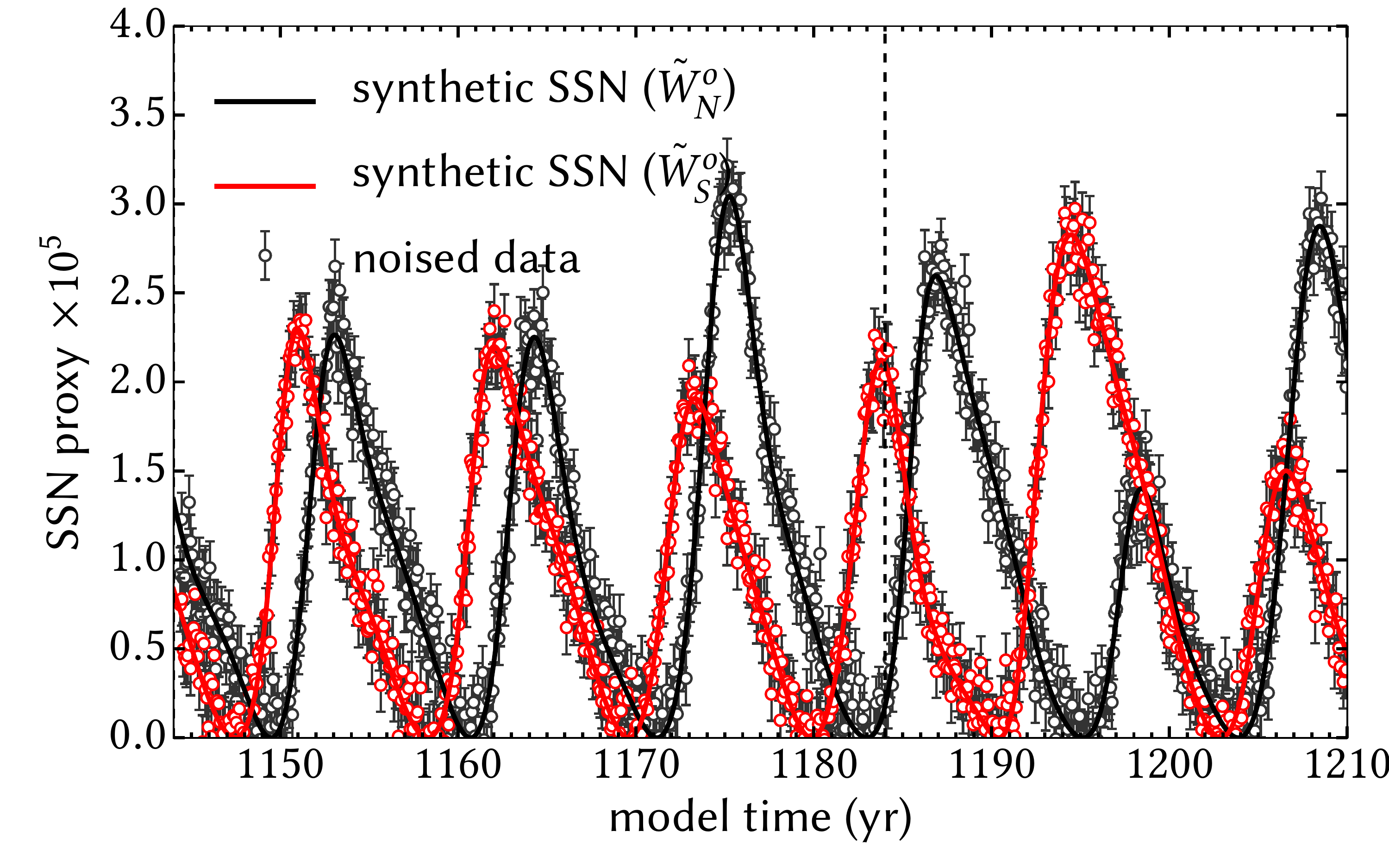}
\caption{Time series of the synthetic sunspot number 
in the Northern (resp. Southern) hemisphere shown in black (resp. red). 
Circles represent the monthly data extracted from these reference time series and 
used for the assimilation. A $10\%$ random error has been added to the reference values 
to generate this data.}
\label{fig:syntheticssn}
\end{figure}

\hcpi{
Since the model does not produce sunspots per se, 
we introduce a proxy for the total sunspot number, in the form of a pseudo-Wolf number $\tilde{W}^o$ defined by
}
\begin{equation}
\tilde{W}^o(t) =  
\left[ 
\int_{\theta=0}^{\theta=\pi} \int_{r=0.70}^{r=0.71} 
B_{\phi}^o(r, \theta, t) r^2 \sin \theta 
\quad 
\mathrm{d}r \mathrm{d} \theta
\right]^2, 
\label{eq:bssn}
\end{equation}
where the superscript $^o$ denotes observations, and the radial coordinate $r$ 
is normalized with the solar radius $R_{\odot}$. 
\hcpi{
We further decompose $\tilde{W}^o$ into its North and South components
}
\begin{equation}
\tilde{W}^o(t) = \tilde{W}_N^o(t)+\tilde{W}_S^o(t), 
\label{eq:bssnsn}
\end{equation}
\hcpi{
in which the North (resp. South) component $\tilde{W}_N^o$ (resp. $\tilde{W}_S^o$) is computed 
by restricting the integration in Eq.~\ref{eq:bssn} to the Northern (resp. Southern) hemisphere. 
In radius, the integral is restricted to a thin layer (between $0.70R_{\odot}$ and $0.71R_{\odot}$) 
where toroidal flux tubes are thought to originate. 
The corresponding pseudo-Wolf numbers are shown in Fig.~\ref{fig:syntheticssn}. 
As the flow applied is equatorially asymmetric, so do the corresponding magnetic proxies. 
Furthermore, there is a clear phase difference between $\tilde{W}_N^o(t)$ and $\tilde{W}_S^o(t)$, 
which suggests symmetric and anti-symmetric dynamo modes as well \citep{Derosa2012ApJ}. 
In these figures, 
note that the \hcpi{$y$-axis} and rightmost dotted lines represent the edges of the 40-year time window
over which we will conduct our assimilation experiments.} \par 
\hcpi{
Synthetic (and noised) time series of $\tilde{W}_N^o(t)$ and $\tilde{W}_S^o(t)$ will 
 constitute one kind of synthetic observations used in our assimilation experiments. The other
 class of data will consist of synthetic (and noised) maps of the line-of-sight component 
 of the magnetic field at the model surface, $B_{los}^o$, defined as
}
\begin{equation}
\begin{split}
B_{los}^o (\theta,t) &= B_r^o (r=1, \theta,t) \sin \theta \\
&=\mathbf{e}_r \cdot \nabla \times (A_{\phi} \mathbf{e}_{\phi}) \\
&= (\cos \theta + \sin \theta \partial_\theta ) A_\phi^o (r=1, \theta, t).
\end{split}
\label{eq:blos}
\end{equation}
\hcpi{
The level of noise should be consistent with that of the observations of the Sun.
We estimate the noise of the surface magnetic field
from the ratio of its coefficient of monopole to the coefficient of dipole
component of the observed field (the former, theoretically, should be zero for noise free situation).
The data is available at WSO,
and the ratio is  $\sim 10\%$.
}
\hcpi{
For the modeled sunspot number proxy $\tilde{W}_N, \tilde{W}_S$,
we refer to real sunspot number data,
the average uncertainty of the data is about $10\%$ of the root mean square of the whole time series
(estimated from sunspot series provided by Solar Influences Data Analysis Center (SIDC)).}
Therefore, we add $10\%$ noise (with respect to the root mean square of the observables)
to the synthetic data $B_{los}^o$ and $\tilde{W}_N^o, \tilde{W}_S^o$ for our numerical experiment. 
Note that this $10\%$ added noise differs from the stochastic forcing $A_i$ of the meridional flow, 
it comes in addition to the fluctuating time series.

\subsection{Assimilation setting}
\label{subsec:assimsetup}
In this section we describe the data assimilation procedure that we have developed to minimize 
the misfit between synthetic observations and magnetic trajectories of the dynamo model, 
 by estimating the meridional circulation and the initial conditions 
which give an optimal fit to the data. 
We also present some technical details in Appendix \ref{sec:asscov}.
The meridional circulation depends on time, 
and from a study of the observed surface flow \citep{Ulrich2010, BasuMC23cycle10, Komm2015SolarPhys} 
the temporal variability is of the order of one year.
Therefore we use an assimilation window of width one year, 
and we will assimilate data for 40 consecutive years. \hcpii{We should stress at this stage
 that within this one-year window, the flow is steady. It can vary from one window to the next, if
 data demand it; the flow is therefore mathematically speaking piecewise constant, over intervals of 
  constant width one~year. 
 }
We choose a course of 40 years because in the future we intend to apply our method 
to invert the magnetic field on the solar surface, 
using the systematic, daily observations the field on the solar disk 
from  WSO (which are digitalized and available from 1975 onward). 
\par 
We include the initial magnetic field of the model in the control parameters ;  
this is a new property of our method compared to \cite{Hungetal2015ApJ} \hcpi{(Paper I)}. 
In \hcpi{Paper I}, the initial condition was approximated 
by the magnetic configuration of a dynamo field produced by a model with 
a steady flow. 
This approximation gets worse when the field is based on a time-varying flow, \hcpii{to the point
 where it precludes the success of the assimilation}.
The assimilation model relies on solving the flux transport model as an initial value problem, 
so we need a better control of the initial conditions. As a result, we add it to the 
control vector together with the flow.
We then express schematically the control vector $\mathbf{x}$ as:
\begin{equation}
\mathbf{x}_n=(\mathrm{x}_{n,IC}, \mathrm{x}_{n,MC})^T.
\label{eq:xvect1}
\end{equation}
Here subscript $1 \le n \le 40$ denotes the step of the assimilation window. 
The component $\mathrm{x}_{n,IC}$ represents the initial conditions in the parameter space, 
and $\mathrm{x}_{n,MC}$ is the meridional circulation, 
which is represented by $c_1$ and $c_2$. 
For our current study there will be 2 coefficients 
representing 2 different structures of flow.
For the initial condition state vector $\mathrm{x}_{n,IC}$, we will discuss below that 
we restrict its dimension to $m=20$, and further justify the consistency between this choice 
and the results in Appendix \ref{sec:asscov}.
\par
The initial conditions for the assimilation model 
$A_{\phi}(r,\theta,t_s)$ and $B_{\phi}(r,\theta,t_s)$, 
where $t_s$ is the beginning of the assimilation window, 
are defined on the grid of $n_r \times n_{\theta}=129\times129$ points. 
However, if we represent the initial condition pointwise in the parameter space,
the number of parameters ($2n_rn_{\theta}\sim32000$) will be too large compared to the number of observations, 
which results in over-fitting. For the latter, 
let $N^o_t$, $N^o_{\theta}$ be the number of sampling in time and latitude respectively,
and the total number of observations $N^o=N^o_{\theta}N^o_t+2N^o_t$. 
(The second term on the right hand side corresponds to \hcpi{$\tilde{W}_N^o$ and $\tilde{W}_S^o$} 
(if they are included as observations).)
At the same grid size, 
an assimilation window of 1 year (sampled on a monthly basis, i.e. $N^o_t=12$) of the surface magnetic field 
(spatial sampling in every latitudinal grid point except the poles $N^o_{\theta}=127$) only gives 
$N^o\sim1500$. 
To stay realistic we do not want an artificially fine sampling in time which 
of course can give a higher $N^o$. 
In practice, sampling frequency of the real magnetic field is, for example, daily in WSO down
to 45 $s$ cadence with HMI onboard SDO satellites \citep{Schou2012SoPh}.  
Latitudinal resolution on real data
also depends on the instrument used. 

\begin{figure}[!ht]
\includegraphics[draft=\draftgraphicx,width=\columnwidth]{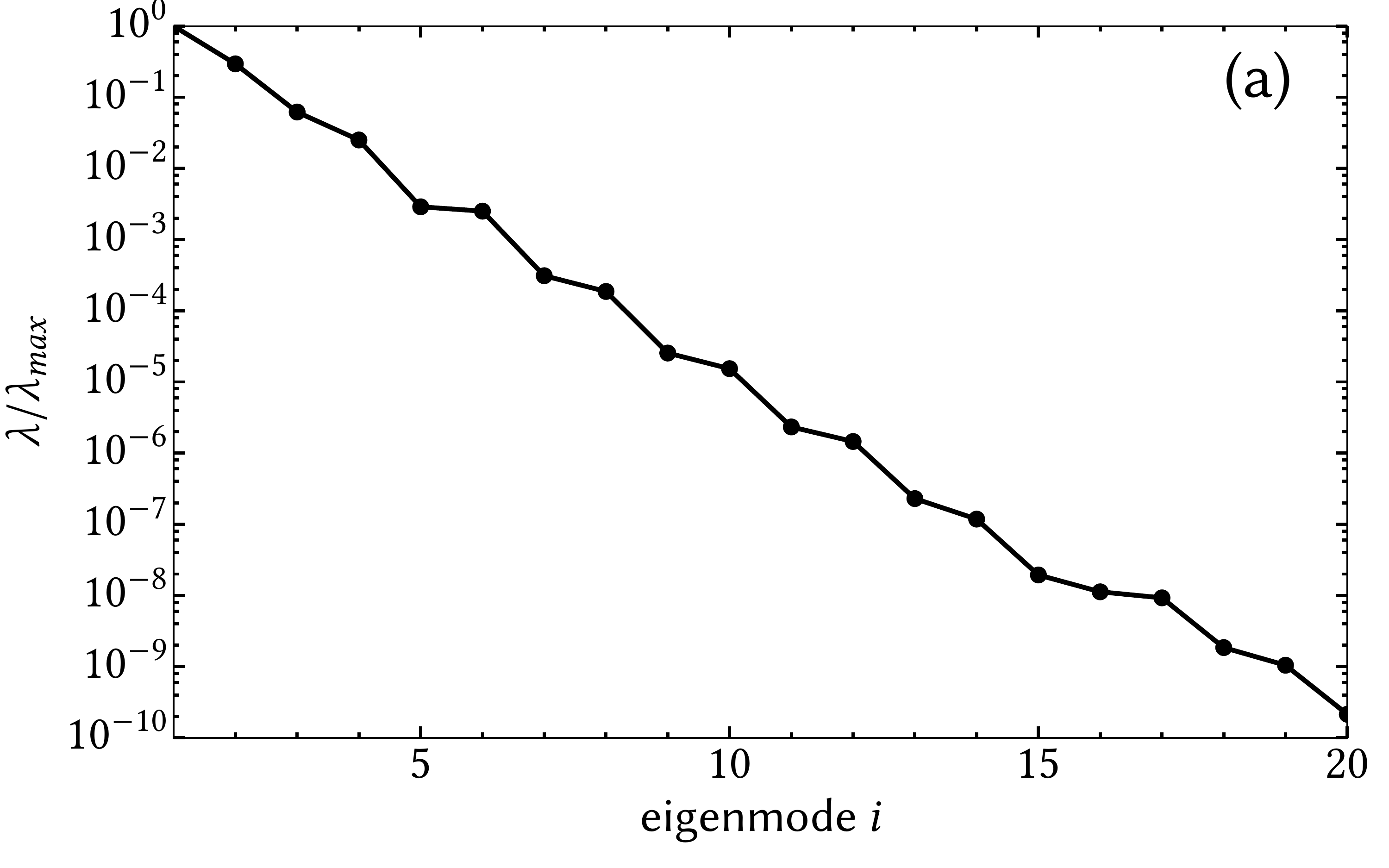}\\
\includegraphics[draft=\draftgraphicx,width=\columnwidth]{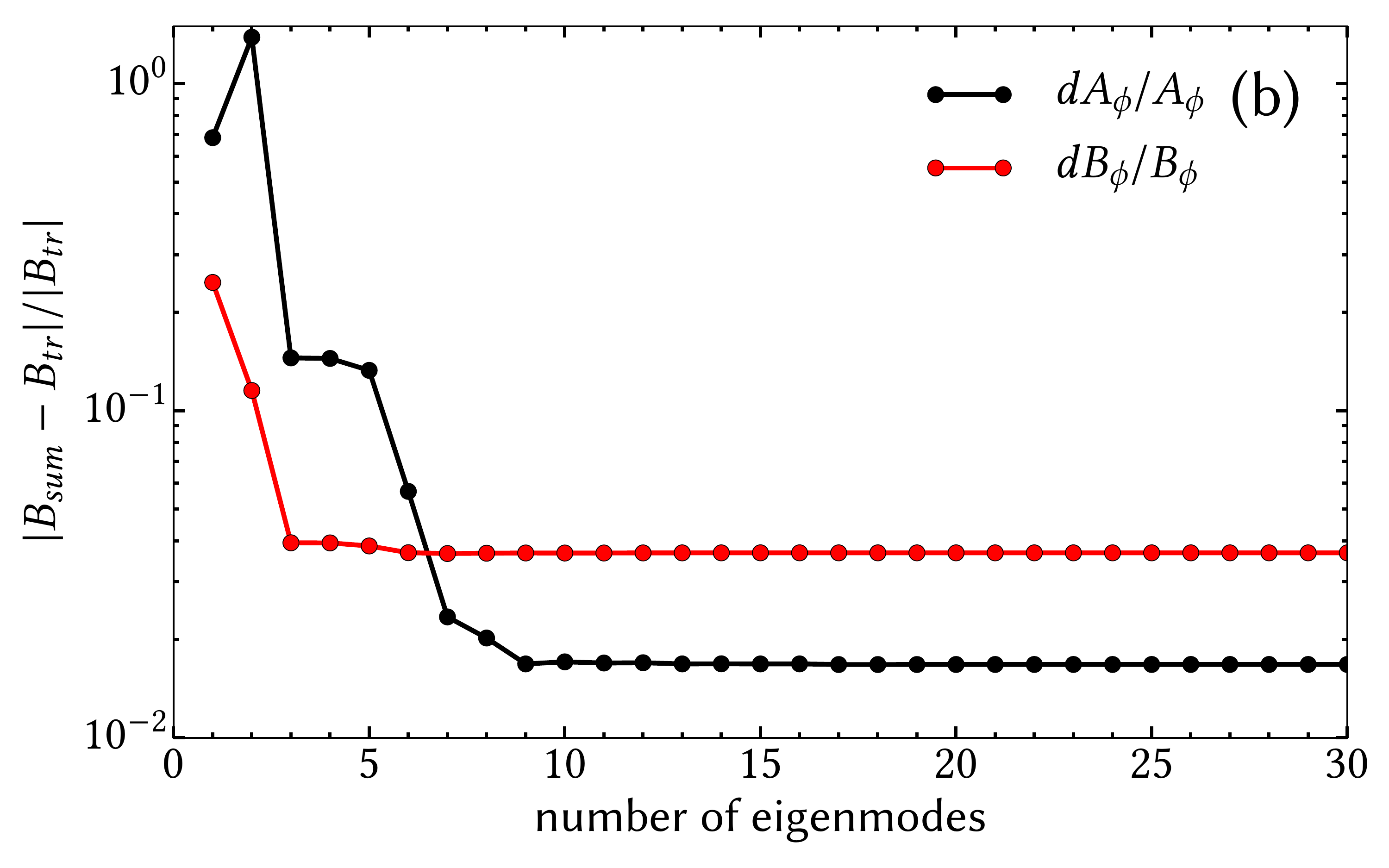}
\caption{(a) Eigenvalue spectrum of the covariance matrix of the dynamo field for a steady unicellular flow. 
\hcpii{The eigenvalues $\lambda$ are normalized with the greatest eigenvalue $\lambda_{max}$ in the plot.} 
(b) Error in the approximation of the magnetic field of the same dynamo field at a particular 
time $t_o=2.91(R_\sun^2/\eta_t)$ as a function of the size of a truncated eigenbasis.
Black: error in the poloidal field. Red: error in the toroidal field.
}
\label{fig:ecovar}
\end{figure}
For instance, the maximum spherical harmonic degree $\ell_{max}$, 
is 60 for WSO maps and about 190 for MDI (note that HMI has 16 times the resolution of MDI) \citep{Scherrer1995SoPh}. 
Therefore, we choose to represent the magnetic field on the meridional plane with 
a truncated set of basis functions rather than pointwise. 
This comes down to constructing the covariance matrix of the dynamo magnetic field $\mathbf{P}$, 
to account for the magnetic variability of the Sun. 
We define and discuss the construction of such a covariance matrix $\mathbf{P}$ in detail in Appendix \ref{sec:asscov}. 
We can then describe our initial magnetic state by retaining only the most prominent eigenvectors of $\mathbf{P}$ as a basis.

Fig. \ref{fig:ecovar} shows the \hcpii{eigenvalue spectrum $\lambda$} and the approximation of the magnetic field driven by 
a simple unicellular meridional flow, with the eigenbasis of its own covariance matrix.
We define the error of approximating the field as
\begin{equation}
 \mathrm{d}X/X = \sqrt{\int_D (X_{approx}-X_{true})^2 \mathrm{d}a \text{\bigg/} \int_D X_{true}^2 \mathrm{d}a},
\label{eq:errorfield}
\end{equation}
where $\mathrm{d}a \sim r \mathrm{d}r \mathrm{d} \theta$,  $X$ is $A_{\phi}$ or $B_{\phi}$, 
$dX/X$ is the error in approximating $A_{\phi, true}$ or $B_{\phi, true}$ with 
$A_{\phi, approx}$ or $B_{\phi, approx}$, respectively. The domain of integration $D$ is the meridional plane.
\par

We can see that we get a good approximation with only $m=20$ basis functions (Fig. \ref{fig:ecovar} (b)), 
so we will limit the number of parameters for the initial condition at $m=20$ under this representation. 
We also update the covariance matrix $\mathbf{P}_n$ with the most recent forecast at the end of each year $n$. 
The dimension of $\mathrm{x}_{n, IC}$ is $20$, together with the 2 parameters in  $\mathrm{x}_{n, MC}$, 
the dimension of the parameter space is $22$, well below $N^o=1500$. 

For the first year of the assimilation window, 
the initial guess for the initial condition ($\mathrm{x}^g_{1,IC}$) 
and meridional flow ($\mathrm{x}^g_{1,MC}$) comes from 
a dynamo model based on a unicellular flow with a magnetic cycle of 22 years (superscript $g$ stands for guess).
Then, for the subsequent data assimilations, the initial guess 
will be the forecast magnetic field and velocity at the end of the previous assimilation.
The former is obtained by evolving the dynamo model for one year with $\mathrm{x}^f_{n-1,IC}$, 
and $\mathrm{x}^f_{n-1,MC}$ (with superscript $f$ stands for forecast), the latter is simply 
$\mathrm{x}^f_{n-1,MC}$.
Within each 1 year window, we estimate the coefficients of the stream function and initial condition 
which give minimal misfit, 
and consequently we obtain an estimate of the time variation of 
the flow profile in Fig. \ref{fig:timevaryflow} by approximating it with a piecewise constant function.

\begin{figure}[!ht]
\includegraphics[draft=\draftgraphicx,angle=-90,width=1.05\columnwidth, angle=90]{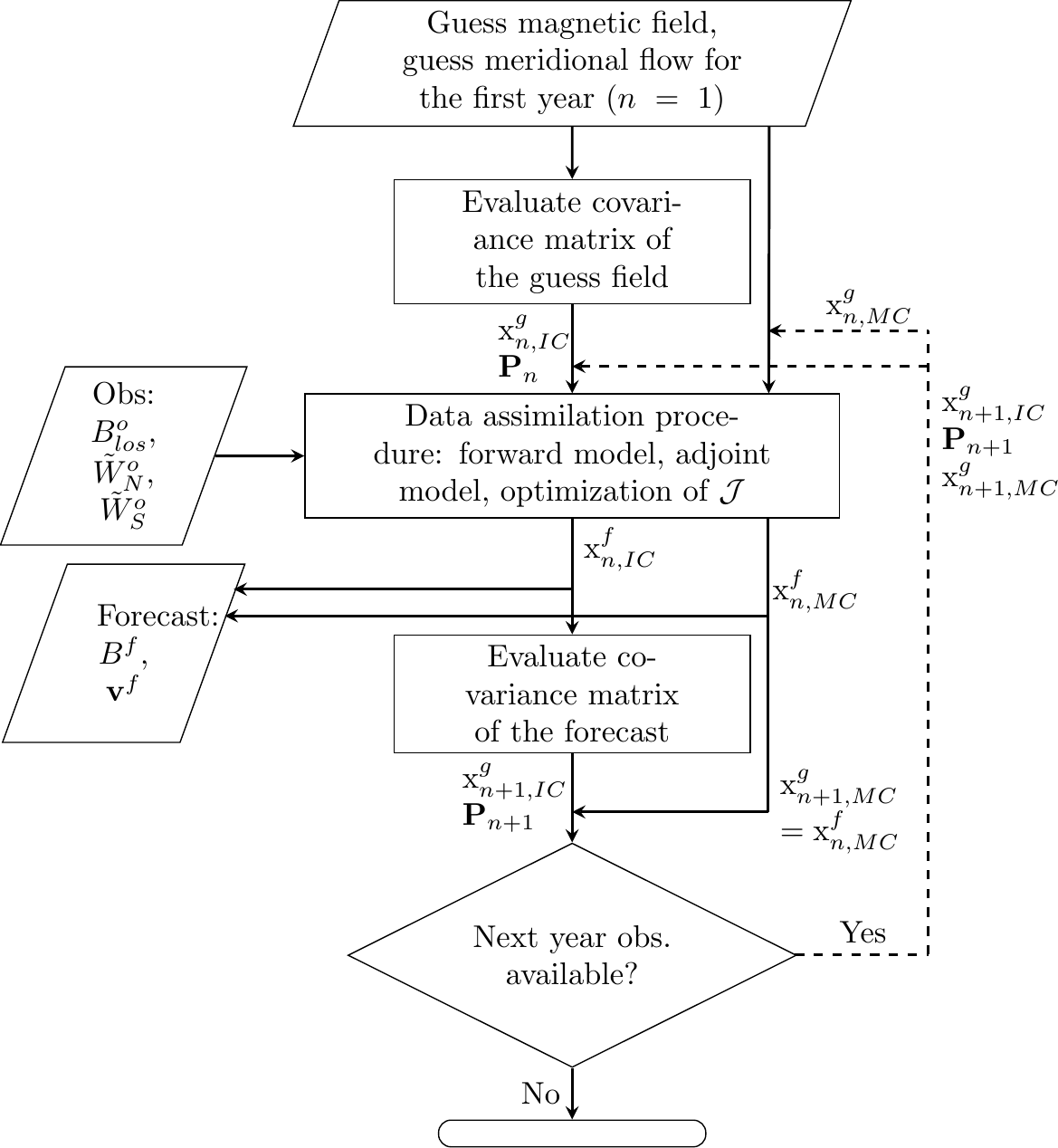}
\caption{A schematic diagram illustrating the data assimilation procedure used in this study. 
Integer subscripts refer to discrete time indices. }
\label{fig:assimschem}
\end{figure}

A schematic view of the procedure is shown in Fig. \ref{fig:assimschem}. 
We start from a guess state representing a dynamo model based on unicellular meridional circulation·
$\mathrm{x}^g_{1,IC}, \mathrm{x}^g_{1,MC}$, and with the input of magnetic observations of the first year, 
we get the forecast state $\mathrm{x}^f_{1,IC}, \mathrm{x}^f_{1,MC}$ from the assimilation procedure. 
Based on the forecast state we can evaluate 
the initial guess state $\mathrm{x}^g_{2,IC}, \mathrm{x}^g_{2,MC}$ of the second year, 
and we repeat the assimilation procedure when new observations are available.

Note that the covariance matrix is evaluated and therefore modified after each year,
so that the projection on the corresponding truncated eigenbasis gives 
a good approximation of the initial condition in each assimilation.
\par

\subsection{Data and objective function}
\label{subsec:choicedata}
We aim to minimize an objective function defined in term of the differences 
between the observations and the model trajectory,
\begin{equation}
\label{eq:objfunc}
\mathcal{J}= \sum_{\alpha} \sum_{i=1}^{N^o_{\alpha,t}} \sum_{j=1}^{N^o_{\alpha,\theta}}  
            \frac{\left[y_{\alpha}({\theta}_j, t_i)-
                        y_{\alpha}^o({\theta}_j,t_i)i\right]^2}
                {\sigma_{\alpha}^2({\theta}_j,t_i)},
\end{equation}
where $\alpha$ denotes the type of magnetic proxy $y$ to be compared. 
The proxies with the superscript $^o$ stand for observations,
and without superscript for the forecast values, 
and \hcpii{$\sigma_{\alpha}$ stands for the uncertainty of the measurement.} 
For each type $\alpha$ we sum the observations over the observation times and latitudes,
$N^o_{\alpha,t}$ and $N^o_{\alpha,\theta}$, respectively. 
\hcpii{Recall that $\mathcal{J}$ is defined over an interval of total duration 1 year.} 
\par
As mentioned above, the synthetic observations used for the experiment 
are the magnetic sunspot proxy (Equation \eqref{eq:bssn}) 
and the surface line of sight magnetic field $B^o_{los}$ (Equation \eqref{eq:blos}). 
Historically, sunspot series given in Wolf number started from 1749,
and daily, continuous and digitalized observations of the surface magnetic field 
of the Sun have become available later. 
Therefore, we first look for the possibility to estimate the (synthetic) 
time varying flow with the assimilation procedure by ingesting 
the modeled synthetic sunspot proxy (Equation \eqref{eq:bssn}) as the only observable. 
This is to investigate the feasibility of estimating the meridional circulation of the Sun since 1749.
However, this would be more difficult as 
SSN is only one value (two for hemispheric SSN) at a particular observation time,
instead of a latitude map provided by $B^o_{los}$. 
We first make this relatively more challenging attempt of 
assimilating the synthetic hemispheric sunspot proxy only,
with an assimilation window of 1 year, 
with various sampling frequencies, from monthly to every 6 days.
We find that in these attempts, 
the estimate of flow in the first year assimilation is not physical, 
as the surface flow is found to be $20$ times higher than the truth. 
This gives an unstable dynamo model for further assimilation after the first year, 
making the algorithm unstable. 
This is because the information contained in the data is not rich enough 
to estimate the meridional circulation as well as the initial magnetic field 
within the assimilation window concerned. 
Moreover, with the sunspot number alone, there is a sign ambiguity 
for the magnetic field. Furthermore, we showed in \hcpi{Paper I} 
that compared with temporal dependence in observations, 
latitudinal dependence is more important for the estimation of internal dynamics. 
\par
To proceed, we can add more information to the pipeline. 
For example, we can add constraints to the optimization procedure based on 
physical knowledge as a background term in the objective function, 
which does not need more observations. 
Or we can add more observations within the assimilation window. 
In this study, we are going to include the characteristics of the butterfly diagram in the observations. 
As a result, we introduce more observations with spatial data distribution, 
in order to help the minimization of the objective function. 
A more effective objective function to be minimized can then be:
\begin{equation}
\label{eq:objj}
\begin{split}
\mathcal{J} &=\sum_{i=1}^{N^o_t}\left\{ \sum_{j=1}^{N^o_{\theta}}              \right.
\frac{[B_{los}(\theta_j,t_i)-B_{los}^o(\theta_j,t_i)]^2}{\sigmalos^2(\theta_j)} \\
            &+\frac{[\tilde{W}_N(t_i)-\tilde{W}_N^o(t_i)]^2}{\sigmassnn^2}            
            \left.   +\frac{[\tilde{W}_S(t_i)-\tilde{W}_S^o(t_i)]^2}{\sigmassns^2}\right\},
\end{split}
\end{equation}
\hcpii{where $\sigmalos$, $\sigmassnn$, and $\sigmassns$ in this study are 
fractions of the root mean squares of the line of sight surface field, 
synthetic sunspot number proxy in the Northern and Southern hemispheres, 
respectively, in order to model the uncertainties of the data, 
i.e. $\sigma_{\alpha}(\theta_j)=\epsilon_{\alpha}\sqrt{<{y_{\alpha}^o(\theta_j)}^2>_t} $, 
where $\epsilon_{\alpha}$ is the level of noise of the species $\alpha$, and $<\cdot>_t$ is averaging over time.
As stated in Sec. \ref{subsec:modeleq}, the noise levels added to $B_{los}^o$ and $\tilde{W}_N^o, \tilde{W}_S^o$ 
for our numerical experiment are of order $10\%$}.

The total number of observations is $N^o=(N^o_{\theta}+2)N^o_t$. 
In our case, within an assimilation window of 1 year, sampling monthly ($N^o_t=12$),
and uniformly in latitude ($N^o_{\theta}=127$), we have $N^o=1548$.
\citep[Here we also tested that for a coarse sampling in latitude, say $N^o_{\theta}=63$, 
we can get similar results and performance. 
For a systematic study of the effect of latitude sampling on the assimilation procedure, 
see][]{Hungetal2015ApJ} \hcpi{(Paper I)}.  \par
The normalized misfit, is defined as
\begin{equation}
\label{eq:Nmisfit}
\mathcal{J}_{norm}=\sqrt{\sum_{\alpha}\frac{\mathcal{J_{\alpha}}}{N_{\alpha}^o}}.
\end{equation}
An optimal fit gives $\mathcal{J}_{norm} \sim 1$, while $\mathcal{J}_{norm} \gg  1$ indicates the misfit is too 
large considering the noise added to the synthetic observations, and $\mathcal{J}_{norm} \ll  1$ implies statistical overfitting. 

\section{Results of assimilation pipeline}
\label{sec:results}
In this section we demonstrate that by assimilating, 
\hcpii{in a sequence of windows of width 1 year,} the synthetic observations 
displayed in Fig. \ref{fig:syntheticfield} and \ref{fig:syntheticssn}, 
we are able to estimate the meridional flow shown in Fig. \ref{fig:timevaryflow}.
We illustrate the data of 40 years under study and the first year of data for 
assimilation in Fig. \ref{fig:syntheticfield} with broken dashed lines. 
We start the assimilation with a unicellular flow as an initial guess for the meridional circulation. 
For the initial condition on the magnetic field components $A_{\phi}$ and $B_{\phi}$ for the first year of the assimilation, 
we conduct 2 trials with 2 different guesses. 
\hcpi{The first guess is a dynamo field based on a unicellular flow ($c_1=1$, $c_2=0$ in Equation~\eqref{eq:modelflow}), 
where the fields have a definite parity about the equator, i.e., symmetric for $A_{\phi}$ and antisymmetric for $B_{\phi}$. 
The second guess is a dynamo field based on a unicellular flow 
but slightly modified with an antisymmetric flow which contributes to 
$1\%$ of the $v_o$ (of the background flow at the surface).
The flow is then slightly asymmetric and so does the corresponding dynamo field. 
The motivation behind the second trial is an
attempt to account for the equatorially asymmetric nature of the synthetic observations 
in Fig. \ref{fig:syntheticfield} and \ref{fig:syntheticssn}.} 
We discuss separately the hind-cast of the data assimilation for 40 years, 
and the ability of the model to forecast beyond the $40^{th}$ year. 
For the latter, 
we estimate the magnetic field 25 years after the latest assimilation, making a total study of 65 years.
For clarity and convenience in discussion, in the following, 
in our figures where a time evolution is shown, $t=0$ corresponds to the time at which we start to ingest observations, 
i.e., $t=0$ at the left broken vertical line in Fig. \ref{fig:syntheticfield}, 
at (model time) year 1144 of the synthetic observations. \par
\hcpi{In the following, the term {\sl dynamical trajectory} refers to the 
time series of the magnetic field in the computational domain, as predicted 
by the numerical dynamo model. The {\sl true}, or {\sl reference} trajectory is 
the one obtained using the combination of control parameters, initial condition
and time-dependent meridional flow used to generate the synthetic data. This {\sl reference} 
trajectory serves 
as a gauge to evaluate the quality of the {\sl assimilated} trajectory. 
The assimilated 
trajectory has an initial magnetic field vector, and an initial meridional flow which 
are not those of the reference trajectory, and the goal of the assimilation 
is precisely to have this trajectory get closer to the true trajectory. 
In contrast, the term {\sl free run} refers to the trajectory obtained, 
starting from this wrong initial set-up, without assimilating any data. }

\subsection{Hindcast by assimilation of the synthetic data}
\label{subsec:Analysis}
In this section we discuss the results of the reconstructed meridional circulation, 
the misfit of data and the estimate of magnetic field when data is available. 
This is possible, as under the basis of numerical experiment, 
the flow driving the dynamo and resulting magnetic field on the meridional plane are known.
\par 

\begin{figure}[!ht]
\includegraphics[draft=\draftgraphicx,width=\columnwidth]{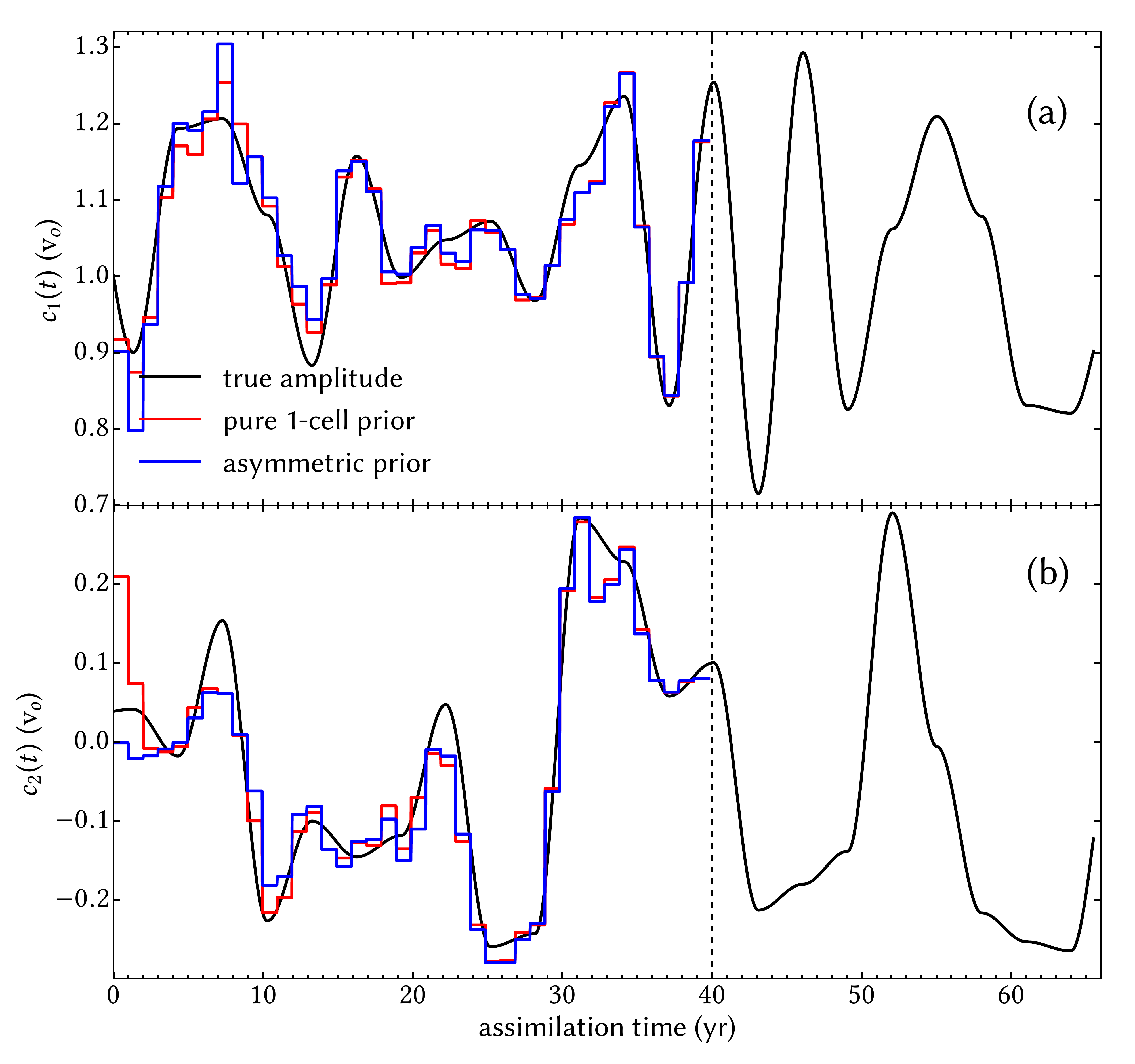}
\caption{(a) Time series of the coefficient of the unicellular component of the stream function.
The reference time series is shown in black. The piecewise constant red (resp. blue) curve 
is the end result of 
the assimilation of synthetic observations starting from a unicellular (resp. asymmetric) 
prior information. 
(b) Same for the coefficient of the antisymmetric component of the stream function. }
\label{fig:recoverflow}
\end{figure}

\begin{figure}[!ht]
\includegraphics[draft=\draftgraphicx,width=0.7\columnwidth, angle=0]{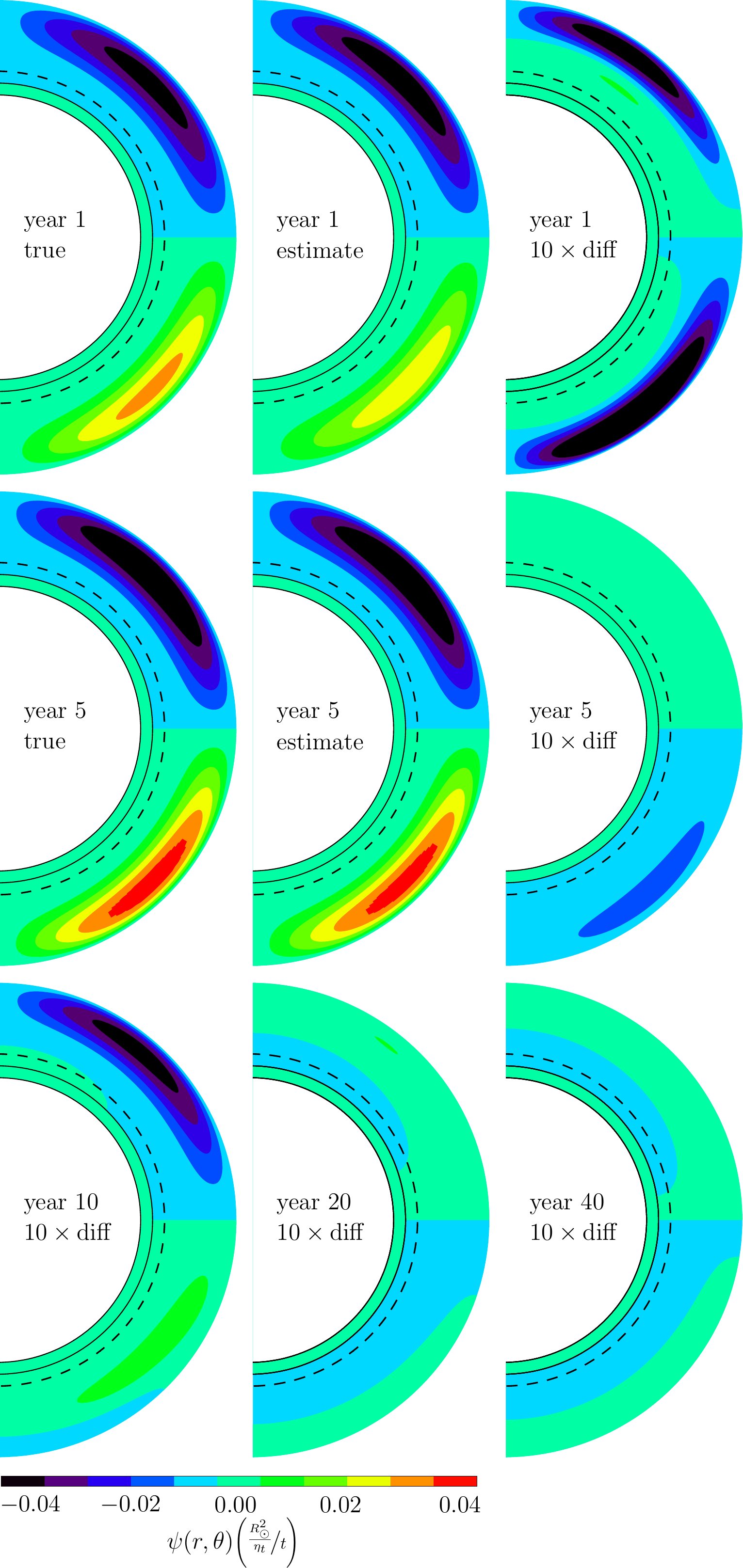}
\caption{The true and estimated stream functions at different epochs during the assimilation 
experiment. Also shown at various epochs is 10 times their differences (estimate-truth). 
}
\label{fig:streammerd}
\end{figure} 

\begin{figure}[!ht]
\includegraphics[draft=\draftgraphicx,width=\columnwidth]{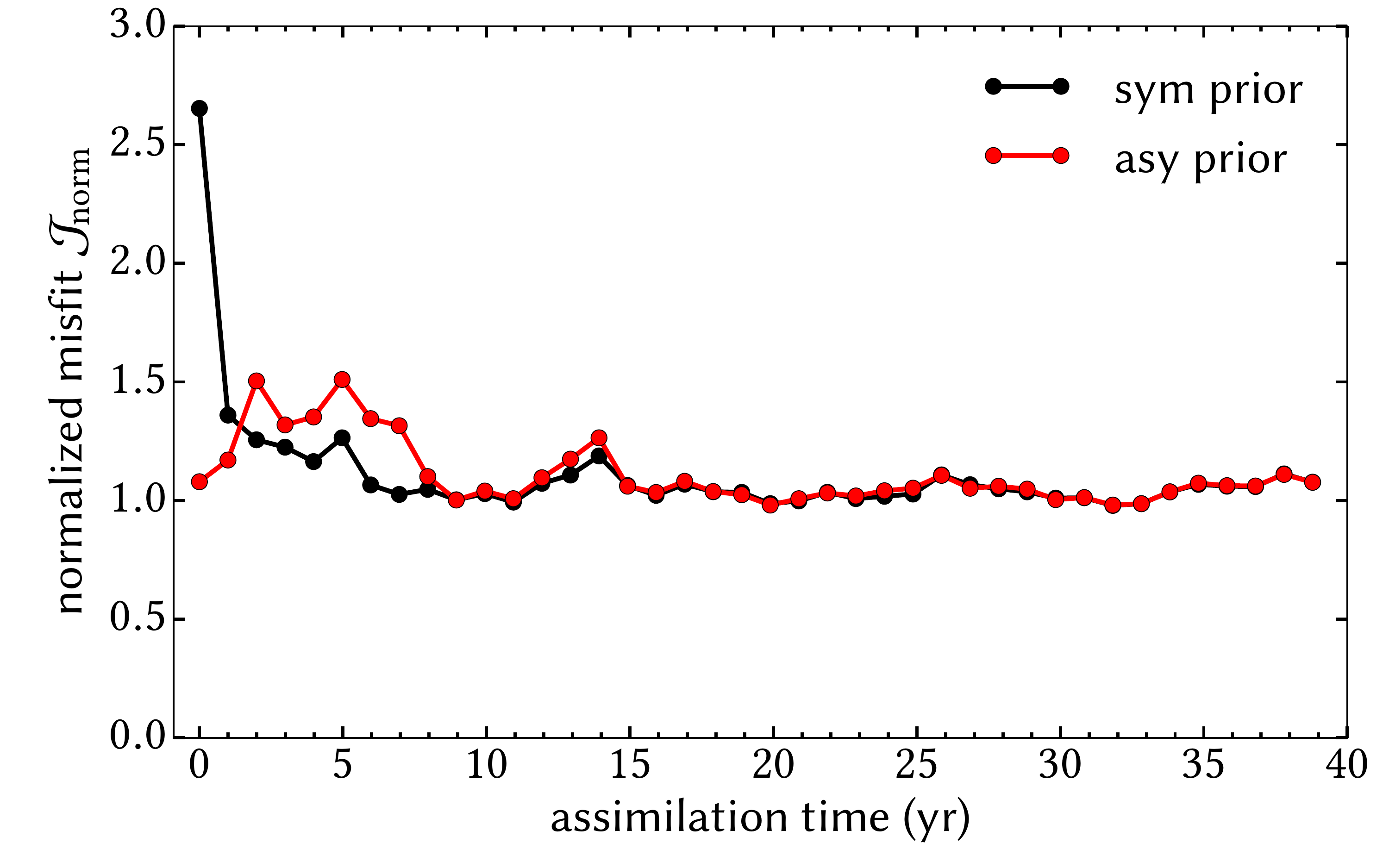}
\caption{Time series of the normalized misfit over the course of the assimilation,
starting either from a unicellular prior (black) or from an asymmetric prior (red). 
See text for details. }
\label{fig:Nmisfitplot}
\end{figure} 

\begin{figure}[!ht]
\includegraphics[draft=\draftgraphicx,width=\columnwidth]{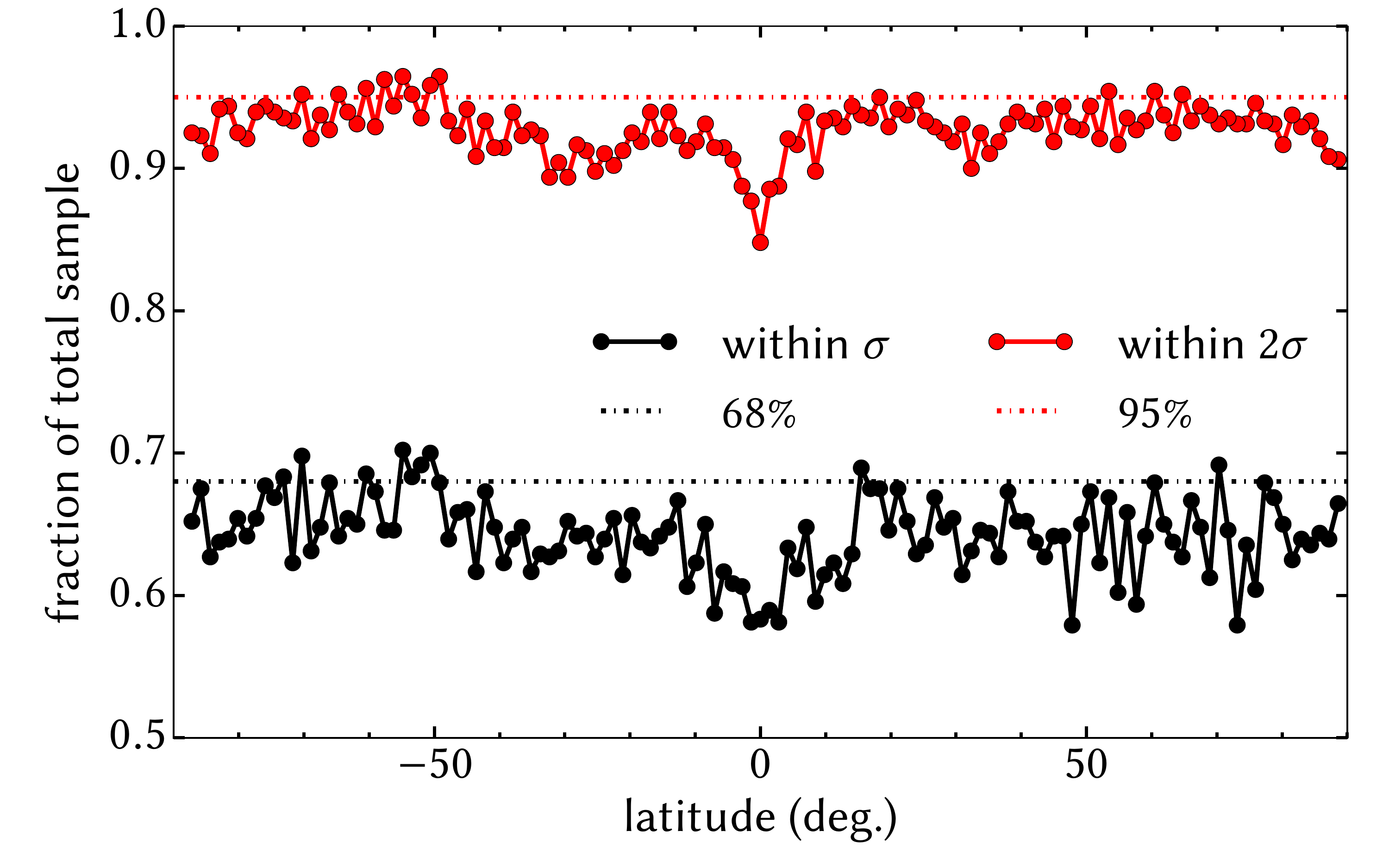}
\caption{\hcpi{
The distribution of the misfit of sampled surface line of sight magnetic field in 40 years of assimilation, 
at different latitudes. 
Black (resp. red) curve: Fraction of estimated surface line of sight magnetic field fall within 
one standard deviation (resp. two standard deviations) from the synthetic observations. 
} }
\label{fig:errlat}
\end{figure} 

\subsubsection{Reconstruction of the time varying flow and minimization of data misfit}
\label{subsubsec:flowmisfit}
We show the estimated coefficients of the stream function in Fig. \ref{fig:recoverflow}. 
By inverting the data, the estimated profiles capture the temporal variation of the stream function reasonably, 
except at the beginning. 

With a unicellular prior for the flow, the difference between the estimate and the truth 
is obvious for the first few years. The synthetic observations are asymmetric about 
the equator as they are based on an equatorially asymmetric true flow. The prior 
for the flow in the first year is symmetric so does the corresponding dynamo field, 
the covariance matrix and eigenbasis of the guess dynamo model. Therefore, 
such a prior cannot take the asymmetry of the observations into account at the onset of the data assimilation. 
As the model in the assimilation technique involves solving an initial value problem where the 
initial conditions are important,  the estimation of the meridional flow is inaccurate. 
However, in data assimilation of subsequent years, the estimated flow starts capturing the asymmetry, 
so does the forecast dynamo field. 
The corresponding updated covariance matrix gives an eigenbasis which can account for asymmetric configuration. 
This shows the ability of the method to adjust the model to give a better approximation to the reality. 
As a result, the estimation of the flow improves starting as soon as the second year. 
In Fig. \ref{fig:streammerd}, we plot the stream functions corresponding to the estimated coefficients (Fig. \ref{fig:recoverflow}),
and (10 times) the differences between the estimate and the truth in some selected years. 
It clearly shows that the error in the estimate of the flow decreases at the beginning.
\par
\begin{figure*}[!ht]
\includegraphics[draft=\draftgraphicx,width=0.95\columnwidth, angle=0]{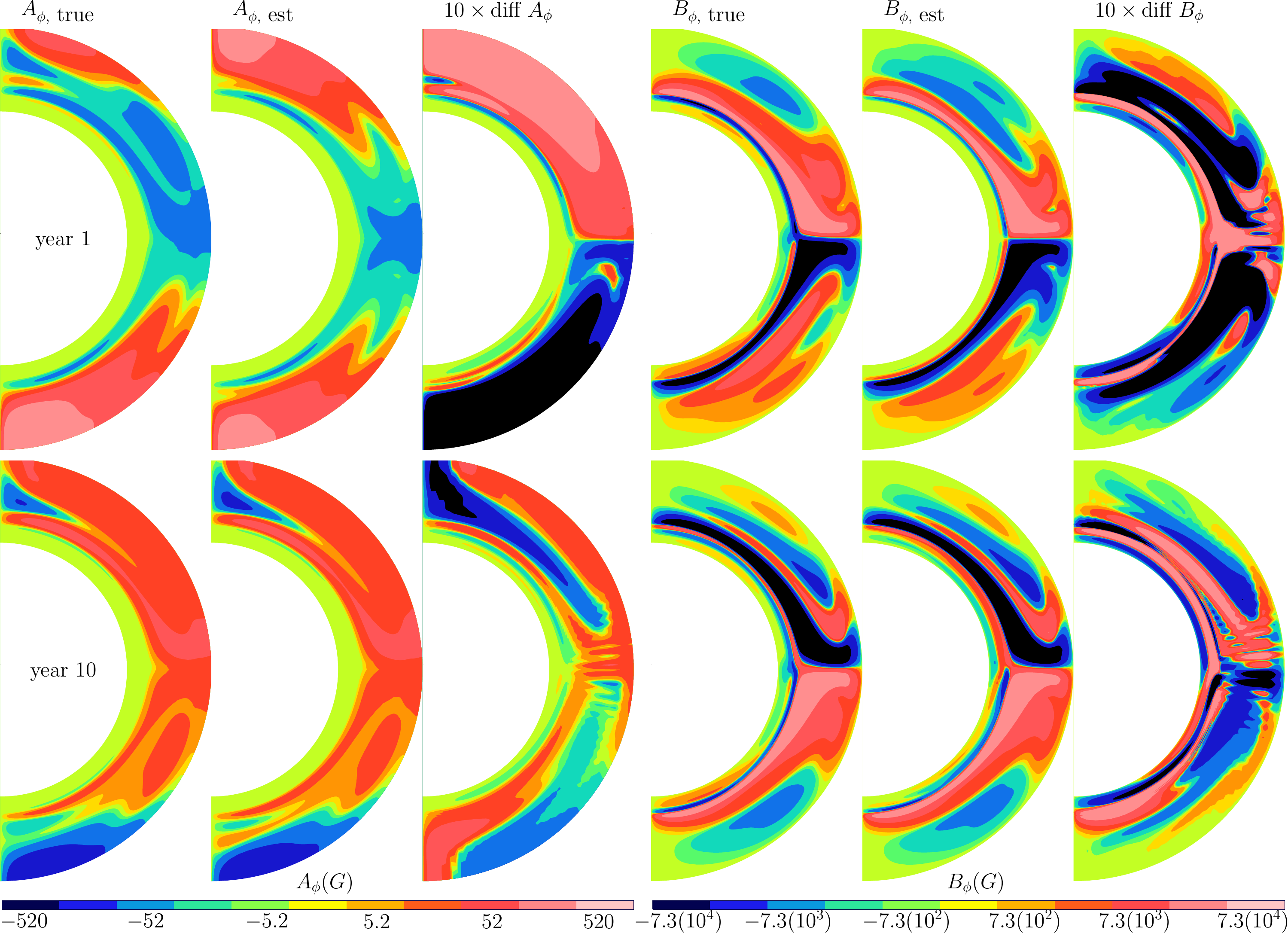}
\caption{Top: meridional plots of the magnetic field at year 1 of the assimilation experiment. 
 From left to right: True poloidal field, estimated poloidal field after assimilation, 
 ten times the differences of these two, true toroidal field, estimated toroidal field after assimilation,
 and ten times the differences of these two. 
 Bottom: same for year 10 of the assimilation. }
\label{fig:icfig}
\end{figure*}

In the trial with a prior based on slightly equatorially asymmetric flow, 
as early as the first year, the covariance matrix is able to account for
the asymmetry of the observations partially. 
Therefore, the estimation of the flow in the first year is better than that obtained using a prior based on 
pure unicellular flow. 
We can also identify such behavior when evaluating the misfit of the synthetic observations in Fig. \ref{fig:Nmisfitplot}. 
Depending on the assumed prior in the first year, 
the normalized misfit is  considerably higher than unity for the first $5$ to $10$ years. 
It converges towards unity after $\sim 10 $ years, 
and remains in very good agreement afterwards.
Also, irrespective of the first year assumed prior, 
the flow reconstructed and the misfit converge to the same value respectively after about $5\sim 10$ years of warm-up time.
\par
The implication here is that the outcome of the assimilation in the first few years
depends highly on initial guess of the initial conditions in the first data assimilation window in this implementation. 
\par
\hcpi{Next, we show in Fig. \ref{fig:errlat} the distribution of the misfit of the surface line of sight magnetic field 
as a function of latitude. As the artificial noise added to generate the synthetic observations 
is normally distributed, theoretically, for optimal fitting, $68\%$ and $95\%$ of the sampled misfit should fall within 
once and twice of the noise level, respectively. The plot shows such a consistency.}

The statistics of the initial guess of the dynamo field determine the 
basis of representation of the initial conditions in the control parameter space. 
An initial guess closer to the truth gives a more complete representation 
and vice versa. This can be improved after assimilation in subsequent years.
Therefore, there is a spin-up time for the assimilation procedure to adapt to the truth, 
but it is reasonably short compared to the interval over which data are available. \par

\newpage
\subsubsection{Estimation of the magnetic field and proxies}
\label{subsubsec:estfield}
\begin{figure}[!ht]
\includegraphics[draft=\draftgraphicx,width=\columnwidth]{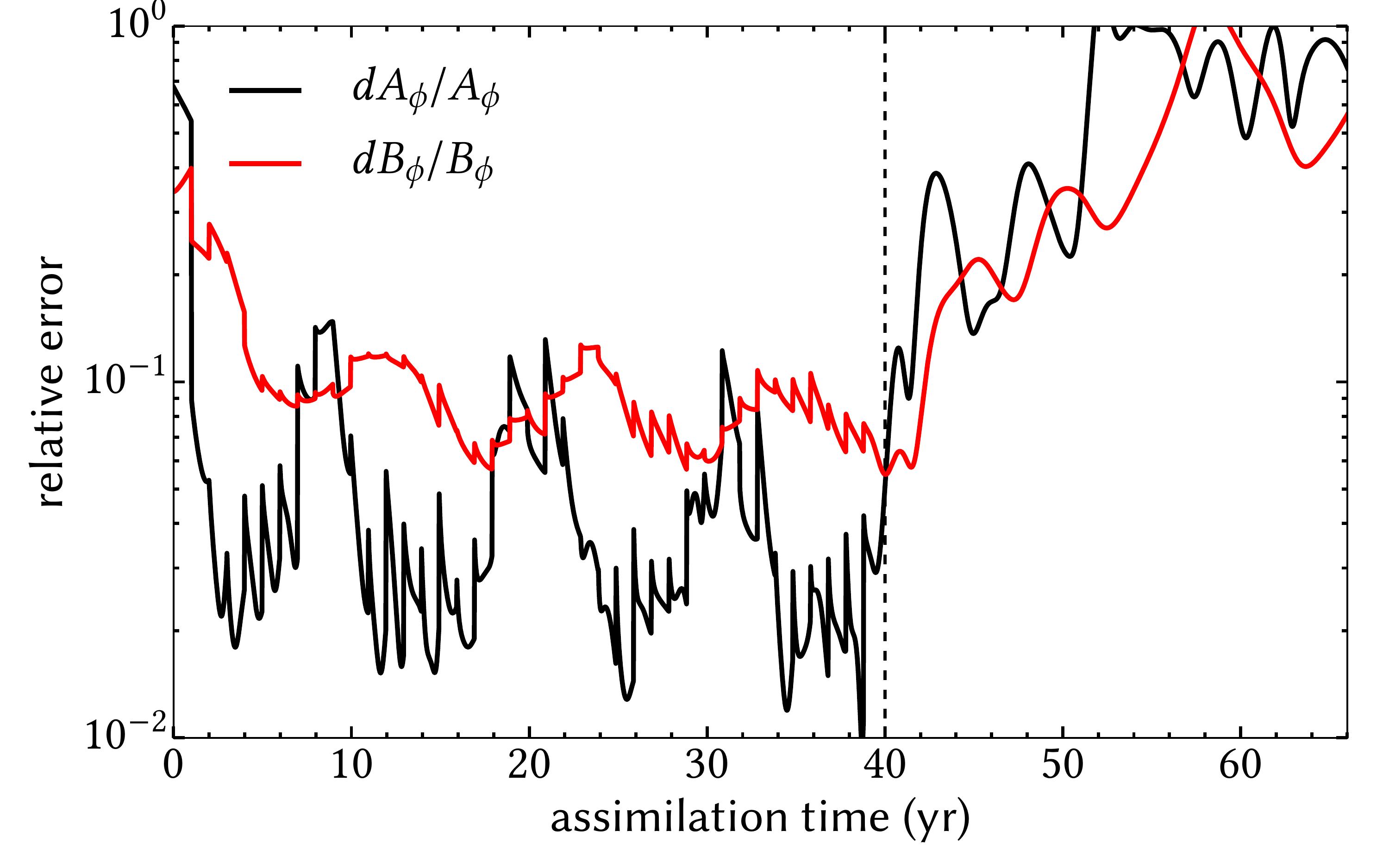}
\caption{Relative difference 
between the magnetic field estimated by data assimilation and the true magnetic field 
versus time, shown in black (resp. red) for the poloidal (resp. toroidal) field.} 
\label{fig:errorB}
\end{figure}

With the estimate of the parameters $\{\mathbf{x}^f_n\}_{1 \le n \le 40}$ (flow, initial condition) 
from the assimilation procedure, we can reconstruct the magnetic field 
and the magnetic proxy (Equation \eqref{eq:bssn}) within the 40 consecutive years. 
We compute the estimated magnetic field $A_{\phi}$ and $B_{\phi}$, and compare with 
the true magnetic field, on the meridional plane (Fig. \ref{fig:icfig}). 
\hcpi{We measure the (relative) error with Equation \eqref{eq:errorfield}. }
We show the difference in Fig. \ref{fig:errorB}. 
The initial guess is based on a unicellular flow.
After the first 5 years of assimilation to capture a dynamo model closer to the truth, 
the relative errors in the estimated field (inside the 40 consecutive years 
of assimilation windows) stay within $10\%$ from the truth. 
This is about the same or slightly more than that of the representation of the magnetic configuration 
introduced in Sec. \ref{subsec:assimsetup} (Fig. \ref{fig:ecovar} (b)). 
\par 

\begin{figure}[!ht]
\includegraphics[draft=\draftgraphicx,width=\columnwidth]{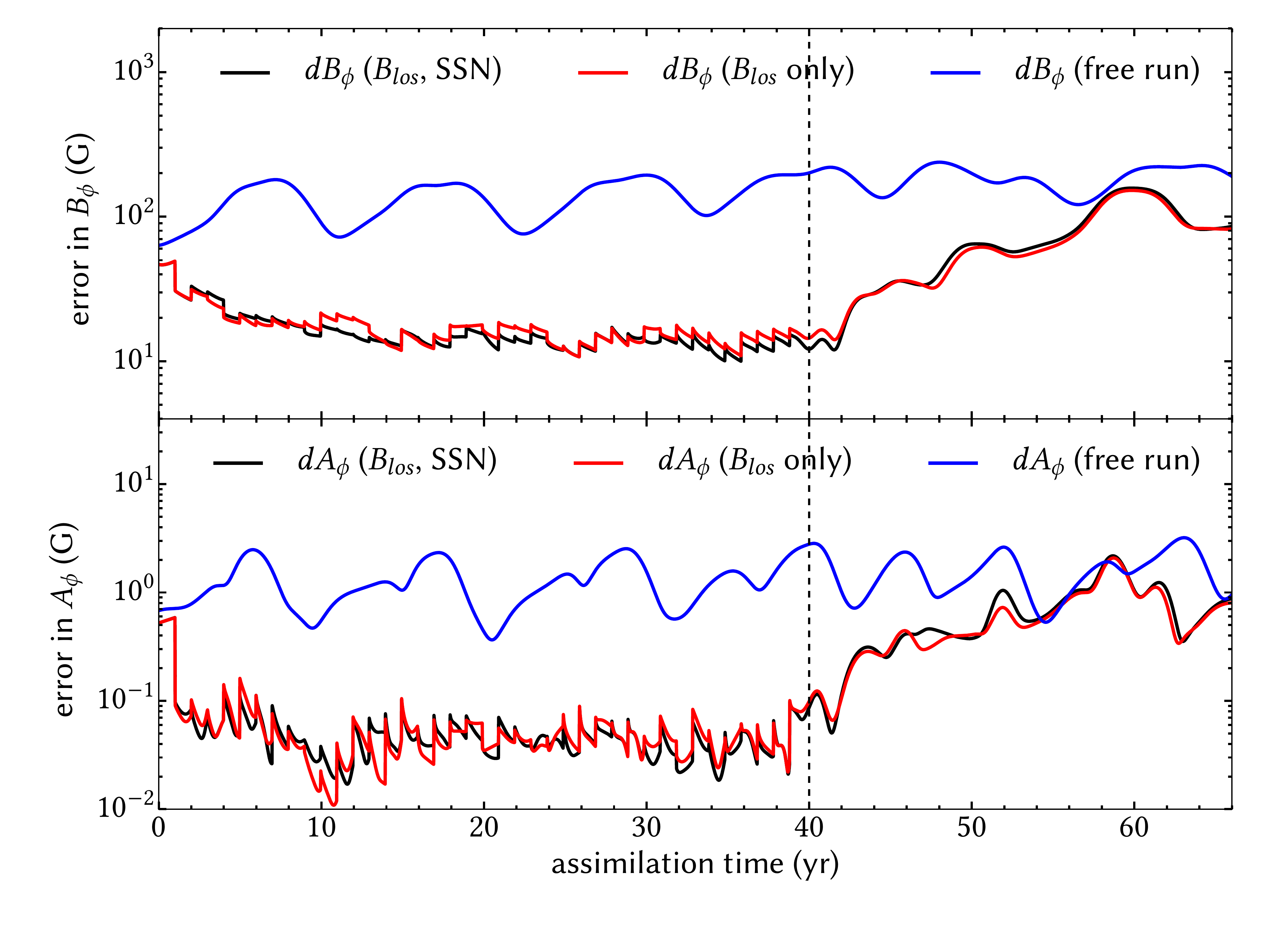}
\caption{Top: Absolute difference between various estimates of the toroidal magnetic field and the 
true magnetic field versus time. Blue: free run of the dynamo model (unconstrained by data). 
Black: data assimilation estimate, with data consisting of magnetic fields in line of sight 
and pseudo sunspot number. Red: data assimilation estimate, with data restricted 
to magnetic fields in line of sight. 
Bottom: Same for the poloidal magnetic field.
}
\label{fig:errorwossn}
\end{figure}
Since the initial magnetic field is in the control parameter space in this assimilation procedure, 
we also show the estimate of the magnetic field on the meridional plane at the first and $10^{th}$
year, the truth and (10 times) their differences, in Fig. \ref{fig:icfig}. 
This also shows that the procedure cannot pick up the asymmetry of the 
field on the first year when the prior is based on a unicellular flow, 
but the asymmetry can be recovered as assimilation time evolves (as shown in year 10).\par
It is believed that the sunspot number is closely related to the toroidal field in the tachocline 
\hcpi{\citep{Parker1993ApJ, Charbonneau1997ApJ, Dikpati99, Choudhuri1995AA}},
so it is important to study the effect of 
data assimilation with the modeled Wolf number $\tilde{W}_N^o$, $\tilde{W}_S^o$ 
on the reconstruction of the magnetic field. 
We compare our reference case with the case 
where only the surface magnetic field is used as observations in Fig. \ref{fig:errorwossn}. 
\hcpi{We also present a free run of a 22-year dynamo model, based on a unicellular meridional flow,} 
and evaluate the difference from our reference model in the same figure for comparison. 
The free run is the situation when there is no data assimilation.
In the presence of the synthetic sunspot-like proxy as observations,
there is only tiny improvement in the estimated toroidal field, 
while the estimated poloidal field is more or less the same. 
This is consistent to the case we showed earlier, that $\tilde{W}_N^o$ and $\tilde{W}_S^o$ alone 
do not give enough information for a reasonable estimate of the state vector. 
The spatial dependence of the observation is important, 
and such a  dependence of the proxy is lost for 
\hcpi{$\tilde{W}_N^o$ and $\tilde{W}_S^o$} 
as it is defined as an integration over latitudes. 
Of course as discussed earlier there are ways to improve the data assimilation algorithm 
based on SSN data only. 
Beyond the 40-year interval of analysis, the error increases when no data is available. 
\par
\begin{figure}[!ht]
\includegraphics[draft=\draftgraphicx,width=\columnwidth]{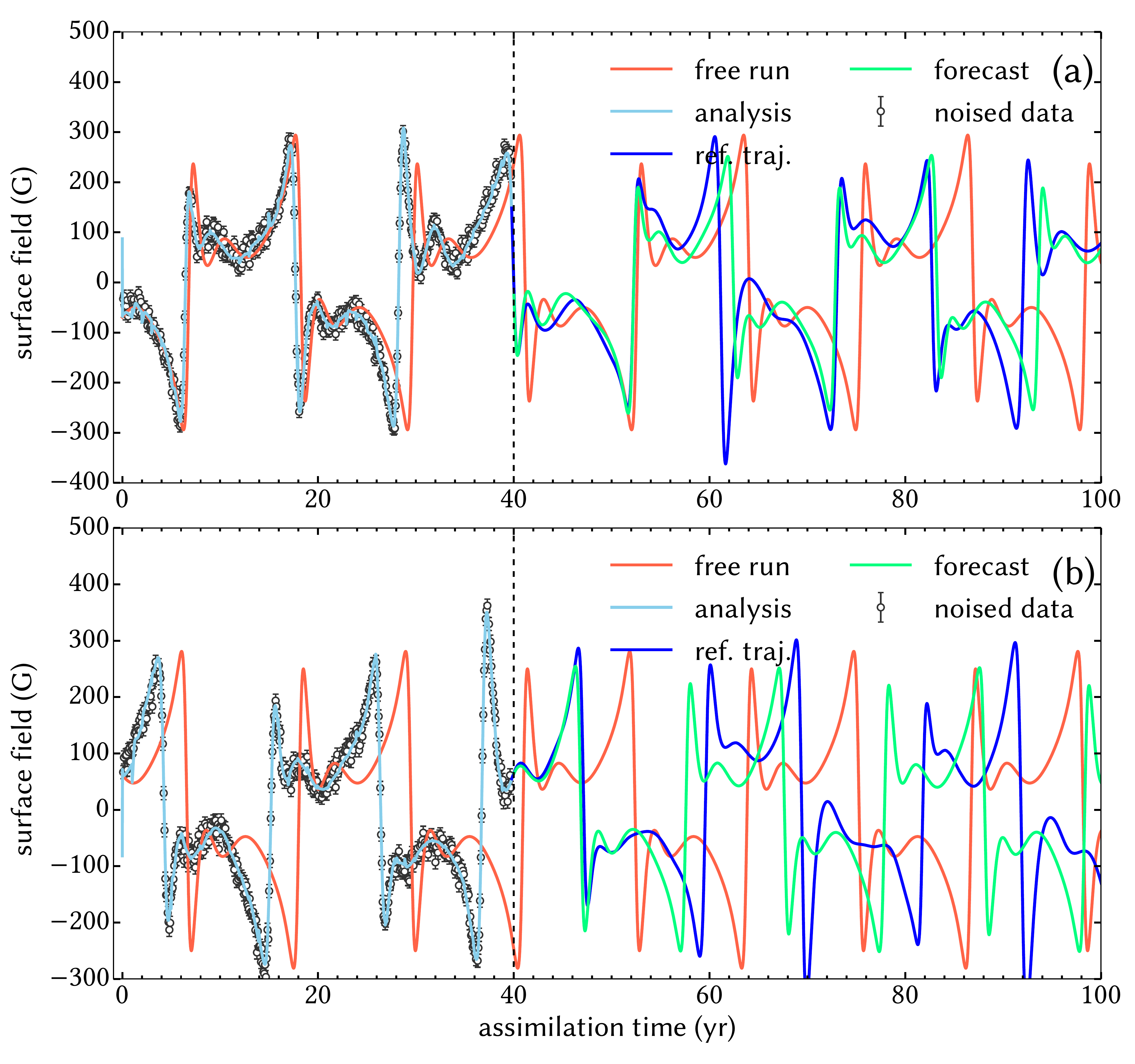}
\caption{Time series of surface magnetic field in line of sight at latitude $20^{\circ}$.
Red: free run of the dynamo model (unconstrained by data). 
Circles: monthly data extracted from the reference time series. 
Blue: reference time series. Light blue: data assimilation estimate. 
Green: time series of the forecast.
(b) Same for the field at latitude $-20^{\circ}$. }
\label{fig:reconsfield}
\end{figure}

Notice there are 2 subtle features about the error in the estimate. 
(i) The tiny and discontinuous rises in error at the beginning of the yearly assimilation windows
shown in Fig. \ref{fig:errorB} are due to an update of 
the truncated eigenbasis of the covariance matrix for each year of assimilation,
which are also within a few $\%$ of the true field. 

(ii) The errors in Fig. \ref{fig:errorB} show a nearly periodic rise and fall for every sunspot cycle. 
As this is an evaluation of the relatively error of the dynamo field, 
and the dynamo field possesses a modulation of cycle $\sim 11$ years (or magnetic cycle of $\sim 22$ years), 
the relative error can be large if the dynamo field is small. 
We show the absolute error of the estimate in Fig. \ref{fig:errorwossn}, 
in which there is no such periodicity in the error. 
(However, Fig. \ref{fig:errorB} illustrates the size of the error compared with the value of the field, 
which is not illustrated in Fig. \ref{fig:errorwossn}). 
In Fig. \ref{fig:errorwossn} we compare the error with \hcpi{the difference between the true trajectory 
and that of} 
a free dynamo run with a simple unicellular flow, without assimilation. 
Compared with the free run, the error decreases in the first 5 years, 
and then the estimated field stays close to the true field until the end of the 40 years series. 
We clearly see the advantage of assimilating data. \par

\begin{figure}[!ht]
\includegraphics[draft=\draftgraphicx,width=\columnwidth]{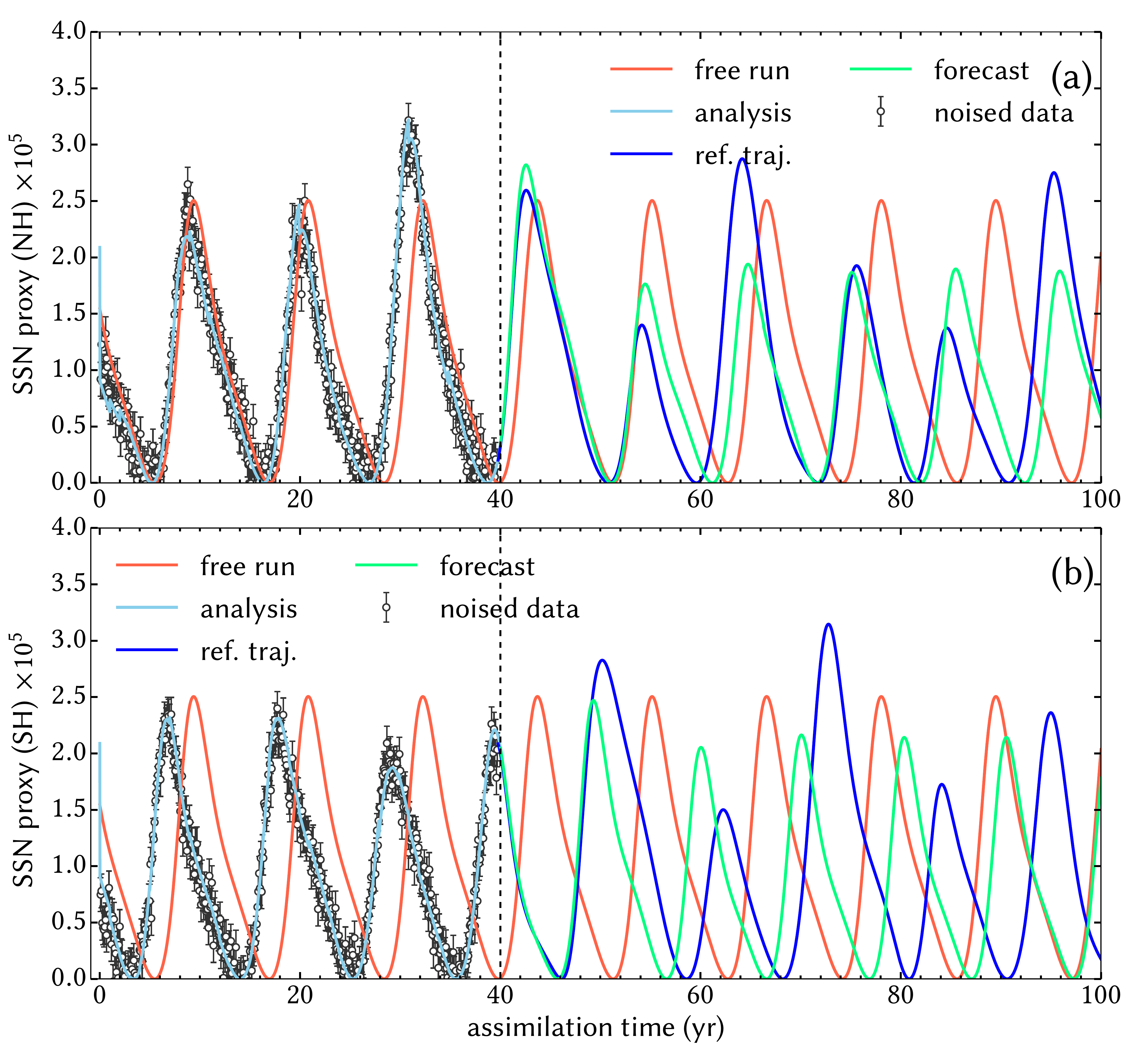}
\caption{(a) Time series of the synthetic sunspot number at the Northern hemisphere. 
Red: free run of the dynamo model (unconstrained by data). 
Circles: monthly data extracted from the reference time series.
Blue: reference time series. Light blue: data assimilation estimate. 
Green: time series of the forecast. 
(b) Same for the synthetic sunspot number at the Southern hemisphere. }

\label{fig:reconsssn}
\end{figure}

\hcpi{Furthermore, we show the fitting of surface magnetic field at latitude \hcpi{$\pm 20^{\circ}$},
in Fig.~\ref{fig:reconsfield}. }
We also show the free run trajectory based on a unicellular flow as reference. 
The synthetic observations are based on an equatorially asymmetric flow, 
so the observations are asymmetric about the equator. 
As a result, we clearly see that the free run quickly goes out of track.
We also note that the free run trajectory based on the symmetric unicellular flow only fits the observations 
reasonably in one hemisphere but not in the other 
(in this case it gets close to the data in the Northern hemisphere.) 
With the assimilation procedure, taking into account the monthly observations \hcpii{each year},
the estimated surface magnetic field reconstructed from the forecast flow, 
gives a smaller misfit in both hemispheres, and clearly the asymmetry is accounted for. 
During the first few years, the misfit is slightly higher than later years,
as the prior is unicellular flow, it takes time for the procedure to adapt to the asymmetry.
Similar results are also observed for the reconstruction of the modeled 
hemispheric sunspot proxy \hcpi{$\tilde{W}_N^o$ and $\tilde{W}_S^o$, defined from the estimated 
toroidal field at the tachocline}, in Fig. \ref{fig:reconsssn}. \par
So we can conclude that our method is robust and able to reconstruct complex, 
possibly asymmetric internal flows from observations of surface magnetic field, 
and yields good agreement with activity in both hemispheres. 
We can now test how well it performs for forecasting. 

\newpage
\subsection{Forecast of the magnetic field and proxies beyond the assimilation window}
\label{subsec:prediction}
In this section we discuss the \hcpii{predictive capability} of the procedure based on this flux transport model.
We estimate the magnetic field beyond the 40 years of assimilation, i.e., without assimilation,
by evolving the dynamo model in time, based on the forecast 
magnetic field and the flow at the end of the $40^{th}$ year. \par
We show in Fig. \ref{fig:errorwossn} the difference between the true field 
and the field obtained from the model beyond 40 years of assimilation.
The error starts to grow for $10 \sim 20$ years but remains smaller than that of the free run. 
After that, it saturates and the 
magnitude of the error is of the same order or slightly lower than that of the free run trajectory.
\par

Therefore, if we try to predict the magnetic observations by extrapolating the model based on 
the magnetic field and the flow at the end of the hindcast,
the prediction is reliable within 10 years if we are conservative, and up to 20 years with low confidence level. 
After 20 years, there is essentially no \hcpii{predictive capability} in this experiment.
This is longer than the time scale of the fluctuations, $\tau = 3$ years, 
added to the reference flow to produce the synthetic observations.
The reasons are (i) the modeled flow contains a non-fluctuating part $\overline{\psi}(\posi)$ 
which is also captured during the assimilation process,
(ii) the long term average of the fluctuations is zero, 
so that the assimilation procedure results in recovering 
the long term averaged flow up to a certain extent. 
\par

In Fig. \ref{fig:reconsfield}, and the modeled sunspot-like proxy in Fig. \ref{fig:reconsssn},
we also show the model trajectory after 40 years when no assimilation is performed.
In particular, for the modeled sunspot proxy in Fig. \ref{fig:reconsssn}, 
the trajectory still fits the observations reasonably after the $40^{th}$ years, 
for $1\sim 2$  cycles ($10\sim20$ years).
And then the trajectory diverges from the observations after 20 years, 
but still closer in phase compared with the free run.
We are then confident that our data assimilation model can provide 
improved predictions \hcpii{in each hemisphere} for up to $15\sim 20$ years. 

\subsection{Numerical experiments with synthetic data based on different levels of stochastic fluctuation on the flow}
\label{subsec:fluctlv}
We showed the estimation of the profile of the flow 
by the data assimilation technique using synthetic observations from 
the flux transport dynamo model with $30\%$ fluctuations on the meridional circulation. 
In this section we study the performance of the assimilation algorithm with respect to 
the magnitude of the fluctuations on the meridional flow 
when generating the synthetic observations.
We test the assimilation method with synthetic observations with $10\%$ and $20\%$ fluctuations, 
together with the $30\%$ case illustrated above.
\par

\begin{figure}[!ht]
\includegraphics[draft=\draftgraphicx,width=\columnwidth]{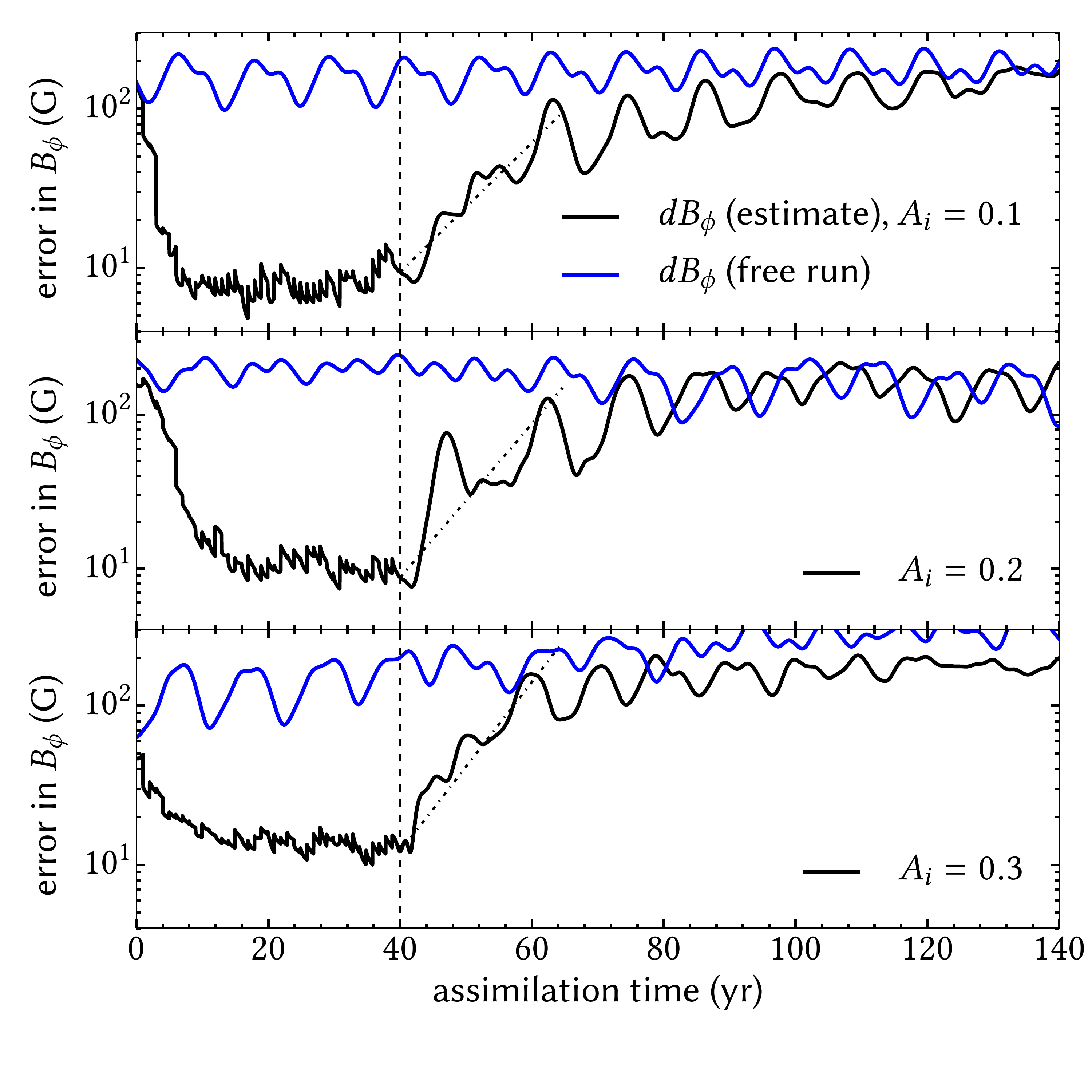}
\caption{ \hcpi{Top: Absolute difference between estimate of the toroidal magnetic field 
and the true magnetic field versus 
time, and where the true magnetic field is driven by meridional flow with $10\%$ fluctuation. 
Blue: free run of the dynamo model (unconstrained by data).   
Black: data assimilation estimate, with data consisting of magnetic fields in line of sight and pseudo sunspot number. 
Middle and bottom: Same for $20\%$ and (resp.) $30\%$ fluctuations in meridional flow.}
}
\label{fig:indifffluct}
\end{figure}

The fluctuations in cycle length of the synthetic observations increase 
with the level of the fluctuation introduced in the meridional circulation.
As illustrated in Fig. \ref{fig:ssncycle}, spread of the cycle length increases 
with the level of fluctuation of the flow.
For the $30\%$ case the range of the distribution 
is comparable to the sunspot cycles, but the lag between the northern and southern 
hemisphere is perhaps too large compared to that the real Sun, so lower 
level of anti-symmetric fluctuations in the flow ($A_2$) is useful to assess. 
\par
To compare the fitting of the synthetic observations of the flow among 3 tests ($10\%$, $20\%$ and $30\%$ fluctuations), 
\hcpi{we show the corresponding 
integrated difference between the estimated and true toroidal field in Fig. \ref{fig:indifffluct}.
As the reference case, both the synthetic sunspot proxy and the surface line of sight magnetic field 
are taken as observations. 
During the 40 years of assimilation,
the absolute difference between the truth and estimate increases with the magnitude 
of the fluctuation of the flow from which the synthetic data is produced.
This is because the approximation of the true flow with step functions become less accurate 
as the fluctuation level (effectively the slope of the profile with respect to time) increases. 
As a result, the difference from the true magnetic configuration is higher.
For the hindcast and forecast process discussed above, i.e., the corresponding results 
shown in Fig. \ref{fig:errorwossn} to Fig. \ref{fig:reconsssn},
the error in estimate of the field and the \hcpii{predictive capability} are of the same order, 
but better for lower fluctuation level $A_i$'s as could be expected (Fig. \ref{fig:indifffluct}). }
\par
Note that in this study, the fluctuation level added to the meridional flow 
imposes little effect on spin-up time.
The time for the \hcpi{integrated difference} to 
decrease and get flattened about unity is $\sim6$ years for the $10\%$ case,
and $\sim10$ years for the $20$ and $30\%$ case.
Therefore, the spin-up time for the assimilation procedures to adapt 
to the truth depends mostly on the initial guess of 
magnetic configuration at the first year of the assimilation pipeline. 
Only a guess closer to the reality can shorten the spin-up time of the procedure.
\par
We conclude from this study that the assimilation procedure is robust 
with respect to the fluctuation level of the time varying flow to be estimated. 
This is important as the latter affects the time variability of the cycle length 
\hcpii{(see Fig. \ref{fig:ssncycle})}. 

\section{Discussion and Summary}
\label{sec:summary}
Our numerical experiment shows the capability of data assimilation 
in estimating the deep meridional circulation of
the Sun using magnetic proxies. 
As a preparation for analyzing real magnetic observations 
and for predicting the solar activity in the future, (in particular cycle 25), 
we adjust the flux transport model to have solar-like properties such as an 11-yr cycle period 
but modulated both in amplitude and frequency, and a time-varying meridional flow 
which may be asymmetric with respect to the equator. 
A stochastic time varying meridional circulation 
produces fluctuations in cycle period and amplitude, 
which make the simulation more solar-like 
compared with a dynamo model with a constant meridional flow, in terms of the irregularities. 
We construct synthetic magnetic proxies, like surface line of sight magnetic field and the sunspot number, 
by relating them to the surface poloidal field and the 
toroidal field in the tachocline computed with the flux transport dynamo model. 
We also add noise to the data and the level of noise is
consistent with the observations from the real Sun ($\sim 10 \%$) 
\hcpii{(recall Sec.~\ref{subsec:modeleq})}. \par
For the data assimilation method, 
we now include the initial conditions of the dynamo model as extra control parameters.
The representation of the initial conditions is based on the statistical covariance of the dynamo model. 
We implement this extension within the corresponding adjoint model, 
such that the resulting framework is capable of 
estimating the meridional flow as well as the magnetic field within the convection zone throughout the
assimilation window. 
We find that the spectrum of the covariance matrix peaks sharply; 
this enables a good approximation of the magnetic configuration
on the meridional plane by projecting it on a truncated eigenbasis (in our test, 20 eigenmodes are taken) 
with the dominant eigenvalues, which facilitates the calculations. \par
We then show that, by ingesting the synthetic (monthly) observations on a yearly basis, 
and within each year applying the 4D-Var assimilation method,
we are able to reconstruct the time varying flow over 40 years of the test period very well. 
The normalized misfit of data, close to unity, indicates an optimal fit in statistical sense.
We also show that the method is robust for synthetic observations 
based on stochastic variations of the flow up to at least $30\%$, 
in terms of reconstruction of the flow and normalized misfit of data. 
By studying the time evolution of differences between the true magnetic field 
from the data and the forecast magnetic field, 
and by further comparing it with a free dynamo run where no data assimilation is done, 
we conclude that in this experiment, the \hcpii{predictive capability} of the method is about 15 to 20 years, 
for the $30\%$ fluctuation in the flow akin to the Sun (exceeding 2 sunspot cycles for lower fluctuation levels). 
Starting from a simple equatorial symmetric dynamo field and unicellular meridional flow, 
the method can give an asymmetric forecast field as well as asymmetric meridional flow, 
hence it is not impaired by symmetry of any sort. 
This is a strength as solar poles are known to reverse with a lag of up to $2$ years 
\citep{Shiota2012ApJ, Derosa2012ApJ}. 
Although there is a spin-up lasting the first $5-10$ years of the assimilation, 
it is short compared to the period over which data are available; its
duration is barely affected by the level of stochastic variation. \par 

\hcpi{
Though we prove the performance of the assimilation procedure with synthetic observations produced by the same flux
transport dynamo model, this is not exactly a twin experiment, since we use step functions to approximate the 
flow in our assimilation model, instead of trying to reconstruct the exact time dependent flow 
that generates the data in Fig. \ref{fig:timevaryflow}. 
}
In generating the synthetic observations, 
the choice of the fluctuation level of the anti-symmetric component ($A_2=0.3$) may seem excessive, 
given the resulting phase difference between both  hemispheres (up to 4 years as opposed to 1-2 years
 for the Sun). 
 Regardless, we show that our pipeline is capable of reconstructing such an asymmetric configuration, while
  disentangling the contribution of both symmetric and antisymmetric flow components to the simulated 
 solar activity. 
In summary,  we are confident that our data assimilation pipeline is robust and a promising tool 
for studying past and future solar activity. 
\par

There are, however, several limitations regarding the model and method used in our study. 
As the flow is perturbed in a stochastic manner, the predictability is limited by the time scale of the stochasticity, 
in our case, 3 years. 
An alternative is to introduce fluctuations in the flow in a non-stochastic manner. 
For example, including the flow as a
dynamical variable of the model which is coupled with the magnetic field nonlinearly, 
which requires a different formulation and closed equation set. 
This more deterministic behavior could actually be more easily captured than a purely random variability.
The long term amplitude modulation, 
such as the Gleissberg cycle, is also absent in the present model, and both can be implemented in future work. 
Regarding the assimilation, we approximate the time varying profile of flow with 
a linear combination of step functions. 
However, we can see that with higher fluctuation on the flow, 
the effect of the slope of the profile becomes important.
The approximation with piecewise constant values will probably give a slightly higher misfit. 
Therefore, a better approximation of the flow in the assimilation routine is necessary. 
Thus, one next important step of improvement is to add the slope of the flow,
i.e., the acceleration, to the control vector of the assimilation framework. 
This will double the number of parameters to represent the stream function.
We show that the method is robust in this relatively hard version of non-constrained numerical experiment. 
On the other hand, it is possible to extend the applicability of the pipeline by introducing physical 
constraints to the framework. 
For example, in the case where we made attempt to hindcast with 
\hcpi{$\tilde{W}_N^o$ and $\tilde{W}_S^o$} alone, 
including more physical information in the form of background term is a possible improvement. 

\hcpii{
At this stage, it may be worthwhile to compare our approach and results with those obtained
 recently by Dikpati and colleagues \citep[][D16 henceforth for the latter]{Dikpati14, Dikpati2016ApJ}. 
D16 carried out 
 a set of numerical experiments using a sequential assimilation method (the EnKF) 
applied to a mean-field dynamo model which resembles closely 
 the one we use in this study. 
The purpose of their proof-of-concept experiments (which rest on synthetic data) is
 to assess the capability of their method to capture the time-dependent behavior
 of the meridional circulation. To that end, they generate a set of synthetic
 observations based on a reference trajectory obtained
 by prescribing a time-dependent meridional circulation. Their meridional 
 circulation has a fixed, one cell per hemisphere configuration, and its time-dependency is restricted to 
 its amplitude. The amplitude has a steady and time-varying part. The time-varying
 part is deterministic, and controlled by a few modes of oscillations with 
 periods of a few years to a decade (see their Figure 1). These deterministic
 oscillations yield fluctuations of about 40\% about the mean (a figure similar
 to the 30\% fluctuations that we generate, in a stochastic fashion though, in this study). 
Their synthetic observations consist of values of the poloidal (at the top of the
 convection zone) or toroidal fields (at the bottom of the convection zone). 
They vary the location and density of observations in their experiments. 
 The true, reference values, are affected by an uncertainty corresponding to
 a noise level of 4\%. This has to be contrasted with observations of the pseudo-number
 of sunspots, and radial induction in the line of sight used here  (affected by 
 relative errors of 10\% throughout our study). 
Dikpati et al convincingly show that by carrying out an analysis every 2 weeks (over the course
 of their 35-yr long experiments) using the EnKF, they can recover the time-dependent 
 amplitude of the meridional flow using an ensemble size of 192 members, each 
 analysis being applied to 10 observations consisting
 of near-surface poloidal fields from low latitudes and tachocline toroidal fields from mid-latitudes. 
Success in retrieving accurately the time-dependent amplitude depends on the locations 
of the available observations (those at high latitude being less valuable).
They also find that a much shorter or longer interval between each update is detrimental
 to the success of the assimilation. A too short an interval (e.g. 5 days) does not allow the system
 to respond dynamically to a change in the flow amplitude, whereas a too large interval between two updates 
 causes the trajectory of the assimilated system to depart excessively from the `true' 
trajectory. 
They do not discuss
  the predictive capability of their system in the study 
(recall that we find in our synthetic setup a practical horizon of predictability of about 15 years).
  Our findings are overall in line with those of D16, in the sense that partial and noised observations
 of a kinematic dynamo with time-dependent flow features can be used to rather accurately
 estimate the time-dependent flow in the bulk of the system (not only where observations 
 are available), by using an interpolation based on a physical model (this is essentially 
 what data assimilation is about).  The differences between their study and ours 
 stand in the assimilation method (sequential vs variational),
and in our estimation of the amplitude and shape of the meridional circulation (as opposed to the amplitude 
alone in D16), in addition to the estimate of the magnetic field.
 In our framework of variation assimilation, we use windows of width 1 year (40 of them for the hindcasting
 part), each of which containing 12 monthly sets of observations. Because we use a 
 similar dynamo model as that used by D16, we also find that observations should be separated by a month
 or so, for the same reasons as those discussed above. 
 With regard to the density of observations, we use more observations at a given
 time (129 versus 10 for D16). Our observations
 are indeed noisier, and, more importantly, the estimation problem that we are 
 looking at is not the same. Our initial set-up for a given window consists of
 the flow properties (2 coefficients) and the initial magnetic field (20 coefficients). 
 In assimilation parlance, our control vector has a size of 22 (recall Sec. \ref{subsec:assimsetup}) whereas, 
in the case of D16, a single parameter (the amplitude) 
 has to be estimated. So it should come as no surprise
 that more observations are needed. 
 In summary, the approach followed by D16 and ours prove capable of estimating
 the time-dependent properties of the meridional circulation in a controlled
 environment (that of a synthetic experiment). We have used synthetic data
 which we think closely resemble the data that we are going to use 
 when dealing with the real (less controlled) problem. 
} \par
To conclude, we presented here an assimilation method to estimate a time varying meridional circulation 
with synthetic magnetic proxies. 
The method is robust with an optimized data fit, and gives a \hcpii{predictive capability} of $1 \sim 2$ sunspot cycles, 
depending on the \hcpii{amplitude of the fluctuating part of the sought flow}. 
 Future developments include (i) analyzing magnetic proxies of the real Sun with the data assimilation method, 
(ii) improving the representation of the meridional circulation in the assimilation framework 
(e.g., by taking into account the acceleration of the fluid), 
and (iii) including physical constraints in the objective function. 
\appendix
\section{The Babcock-Leighton flux transport mean field dynamo model}
\label{subsec:BLmodel}
This section gives a brief description of the flux transport mean field dynamo model, 
i.e., the Babcock-Leighton model, with axisymmetry. This is the model used for the 
assimilation procedure, and to generate synthetic observations for our numerical experiment 
to verify the data assimilation technique.
The model equations are \hcpii{\citep{Dikpati99,JouveMC07,JouveBenchmark08,Hungetal2015ApJ}}: 
\begin{equation}\label{eq:Adyn}
\begin{split}
\partial_t A_{\phi} & =\frac{\eta}{\eta_t}  \left(\nabla^2 - \frac{1}{\varpi^2}\right) A_{\phi}
-Re \frac{\myvect{v}_p}{\varpi}\cdot \nabla (\varpi A_{\phi}) 
  + C_sS(r, \theta, B_{\phi}), 
\end{split}
\end{equation}

\begin{equation}\label{eq:Bdyn}
\begin{split}
\partial_t B_{\phi} = &
\frac{\eta}{\eta_t}  \left(\nabla^2 - \frac{1}{\varpi^2}\right) B_{\phi}·
+\frac{1}{\varpi} \frac{\partial (\varpi B_{\phi})}{\partial r} \frac{\partial (\eta/\eta_t)}{\partial r}·
 -Re \varpi \myvect{v}_p\cdot \nabla \left( \frac{B_{\phi}}{\varpi }\right)·\\
&- Re B_{\phi} \nabla \cdot \myvect{v}_p·
+ C_{\Omega} \varpi \left[ \nabla \times (A_{\phi} \myhat{e_{\phi}}) \right] \cdot \nabla \Omega ,·
\end{split}
\end{equation}·
where $A_{\phi}(\vec{r},t)$ and $B_{\phi}(\vec{r},t)$ are the poloidal potential field 
and the toroidal field respectively. 
$\varpi=r \sin \theta$, and $\myvect{v}_p$ is the poloidal velocity, i.e., the meridional circulation, 
$\Omega$ is  the profile of the differential rotation, 
and $S$ is the source of the poloidal field at the solar surface. 
The domain is $(r,\theta) \in [0.6,1] \times [0,\pi]$. 
The toroidal field $B_{\phi}=0$ at the boundary of the domain, and for $A_{\phi}$, 
we impose the pure radial field approximation at the surface, i.e., $\partial_r (r A_{\phi})=0$ at $r=1$, 
and $A_{\phi}=0$ on all the other boundaries.
The length is normalized with solar radius $R_\sun$, time is normalized with the diffusive time scale $R_\sun^2/{\eta}_t$ 
where ${\eta}_t$ is the envelope diffusivity. 
We introduce 3 dimensionless parameters, namely the Reynolds number based on the meridional flow speed $Re=R_\sun v_o/\eta_t$, 
the strength of the Babcock-Leighton source $C_s=R_\sun s_o/\eta_t$ 
and the strength of the $\Omega$-effect $C_{\Omega}=\Omega_oR_\sun^2/\eta_t$, and $\Omega_o = 2\pi \times 456\mathrm{nHz}$.

We use the same dynamo model as we did in \hcpi{Paper I} 
here except that we have some modifications.
First, we use a slightly more complex resistivity profile, a 2-step profile in radial direction,
\begin{equation}
\label{eq:etaprofile}
\frac{\eta}{\eta_t}=\frac{\eta_c}{\eta_t}+\frac{\eta_m}{2\eta_t}\left[1+\tanh\left(\frac{r-r_{bm}}{d_1}\right)\right]
+\frac{1}{2}\left[1+\tanh\left(\frac{r-r_2}{d_1}\right)\right],
\end{equation}
where $\eta_c=10^9$ cm$^2$~s$^{-1}$, $\eta_m=10^{11}$ cm$^2$~s$^{-1}$, $\eta_t=5\times 10^{11}$ cm$^2$~s$^{-1}$, $r_{bm}=0.72$,
$r_2=0.95$, $d_1=0.016$.
In this resistivity profile, the high diffusion at the surface brings a lower
ratio of radial magnetic field at the pole to that near the equator \citep{HottaYokoyama2010}. 
Second, the meridional circulation is also modified. 
The meridional flow is crucial in this model, it advects the magnetic field poleward at the surface, 
and equatorward deeper in the convection zone when it is unicellular per hemisphere. \par
To obtain dynamo generated magnetic field with fluctuations in period and amplitude 
instead of a constant 22 years and peak amplitude, 
we use a time varying meridional circulation for the model. 
We express the flow in the convection zone as the curl of a stream function:
 \begin{equation}\label{eq:curlvp}
  \myvect{v}_p= \nabla \times (\psi \myhat{e_{\phi}}),
 \end{equation}
 and we expand the stream function as
 \begin{equation}\label{eq:MC}
 \begin{split}
 & \psi (r,\theta,t)= -\frac{2}{\pi}\left(\frac{r-r_{mc}}{1-r_{mc}}\right)^{2.5}(1-r_{mc}) \\
 & \times
   \begin{cases}
      \sum\limits_{k=1}^{m} \sum\limits_{l=1}^{n} d_{k,l}(t) \sin \left[ \frac{k\pi (r-r_{mc})}{1-r_{mc}} \right]
  P_\ell^1 (-\cos \theta) & \text{if }r_{mc} \leq r \leq 1 \\
      0 & \text{if } r_{bot} \leq r<r_{mc},
   \end{cases}
 \end{split}
 \end{equation}
where $P^1_\ell$ are the associated Legendre polynomials of order 1. 
The meridional flow is allowed to penetrate to a radius $r_{mc}=0.65$, 
i.e. slightly below the base of the convection zone located at ${r_c}=0.7$. 
Notice that the radial dependence of the stream function is raised to $(r-r_{mc})^{2.5}$, compared with $(r-r_{mc})^2$ in 
\cite{JouveBenchmark08} and \hcpi{Paper I}. 
This can give a higher ratio of maximum flow $v_{\theta}$ at the surface 
with respect to that of the base of the convection zone, which in turn 
results in a 22-year magnetic cycle dynamo model with a surface flow $\sim 20$ ms$^{-1}$ \citep{Yeates2008}, 
consistent with the observed solar surface flow \citep{Ulrich2010, BasuMC23cycle10, Komm2015SolarPhys}. 
The expansion coefficients $d_{k,l}(t)$ are modulated in time 
so that the flow is time dependent.
Other parameters used in the model include the Reynolds number $Re=310$, $C_s=20$, 
$C_{\Omega}=2.78 \times 10^4$, i.e., $v_o=22.3$ms$^{-1}$, $s_o=1.44$ms$^{-1}$. 
The grid size is $n_r \times n_{th} = 129 \times 129$, and the time step is $10^{-6}$, equivalent to $0.112$ day. 
In the illustrative example of the numerical experiment starting from Sec. \ref{subsec:modeleq}, 
\hcpi{we chose a model flow (Equation \eqref{eq:modelflow}) characterized by $d_{1,2}(t) = 1/3c_1(t)$,
and $d_{2,1}(t) = 0.0865c_2(t)$, $d_{2,3}(t) = 0.130c_2(t)$ [$d_{k,l}=0$ for other $(k,l)$'s]. } 
Of course, model based on stream functions defined by different combination of $d_{k,l}$'s can be investigated.

\section{Assimilation procedure and representation of initial conditions in the parameter space}
\label{sec:asscov}
We present here the technical details of incorporating the initial magnetic field of the dynamo model 
to the control parameter space as a reference. \par
The initial conditions for the assimilation model are the magnetic potential of the poloidal field 
and the toroidal magnetic field on the meridional plane at the beginning of an assimilation window, 
i.e., $A_{\phi}(r,\theta, t_s)$ and $B_{\phi}(r,\theta, t_s)$, respectively. 
To extend the parameter space in the present 4D-Var framework, the initial conditions become part of 
the implicit  dependences of the objective function. \par
As mentioned in Sec. \ref{subsec:assimsetup}, we need a representation of 
$A_{\phi}(r,\theta, t_s)$ and $B_{\phi}(r,\theta, t_s)$ in the parameter space 
such that the associated dimension is small compared with $N^o \sim 1500$.
To address this problem, we represent the magnetic field on the meridional plane with 
a truncated set of eigenbasis of the covariance matrix of a dynamo field trajectory. 
We find that for a magnetic trajectory from the flux transport dynamo model 
$A_{\phi}(r,\theta,t)$ and $B_{\phi}(r,\theta,t)$, if we calculate the covariance matrix over a long time 
(which covers the 22-years period of the magnetic cycle), the magnetic field at any time in the trajectory 
$A_{\phi}(r,\theta,t_o)$, $B_{\phi}(r,\theta,t_o)$ can be approximated effectively with a linear combination 
of only the first few eigenvectors of the covariance matrix with leading eigenvalues. 
We define the field column vector 
\begin{equation}
\label{eq:statvect}
\begin{split}
\mathbf{y}(t)= &[A_{1,1}(t),..,A_{n_r,1}(t),A_{1,2}(t),..,A_{i,j}(t),..,A_{n_r,n_{\theta}}(t),  \\ 
               &B_{1,1}(t),..,B_{n_r,1}(t),B_{1,2}(t),..,B_{i,j}(t),..,B_{n_r,n_{\theta}}(t)]^T, 
\end{split}
\end{equation}
where  $X_{i,j}(t)=X(r_i,\theta_j,t)$ with $X$ be $A_{\phi}$ or $B_{\phi}$. 
$r_i,\theta_j$ are the spatial grid points of the magnetic field, so the size of the vector is $2n_r n_{\theta}$, 
with $n_r$, $n_{\theta}$ 
being the grid size in radial and polar direction in the coordinate space respectively. The covariance matrix $\mathbf{P}$ 
about a particular time $t_o$ is defined as
\begin{equation}
\label{eq:covar}
P_{k,l}(t_o)=\overline{ [\mathbf{y}-\mathbf{y}(t_o)]_k[\mathbf{y}-\mathbf{y}(t_o)]_l^T },
\end{equation}
where the over-bar denotes averaging over time, in our case, two magnetic cycles. 
Notice that the indices $k,l$ are the indices of the field vector and 
the covariance matrix, with $1 \le k,l \le 2n_rn_{\theta}$.
The diagonal entries of $\mathbf{P}$ are the variances of $A_{\phi}$ and $B_{\phi}$ at each grid point respectively. 
The off diagonal entries, depending on the indices, are the covariances of $A_{\phi}$ ($B_{\phi}$) between any 2 different grid points,
or the covariances between $A_{\phi}$ and $B_{\phi}$ at any 2 grid points. It measures the auto-correlations of $A_{\phi}$ and $B_{\phi}$,
and also the correlation between  $A_{\phi}$ and $B_{\phi}$.
We diagonalize the matrix, project $\mathbf{y}(t_o)$ on the eigenbasis and approximate 
$\mathbf{y}(t_o)$ in a truncated linear combination of the eigenvectors:
\begin{equation}
\label{eq:expansion}
\mathbf{y}(t_o) \sim \sum_{i=1}^m [\mathbf{w}_i^T\mathbf{y} (t_o)]  \mathbf{w}_i,
\end{equation}
where $\{\mathbf{w}_i\}$ is the eigenbasis of $\mathbf{P}$, 
with the corresponding eigenvalues $\lambda_1 \ge \lambda_2 \ge ...\ge \lambda_m$, 
$m$ is the number of basis vector used in the approximation. 
Notice that the covariance matrix is positive definite and symmetric by definition, 
thus the corresponding eigenbasis is orthonormal. 
\par

\begin{figure}[!ht]
\includegraphics[draft=\draftgraphicx,width=0.9\columnwidth,angle=0]{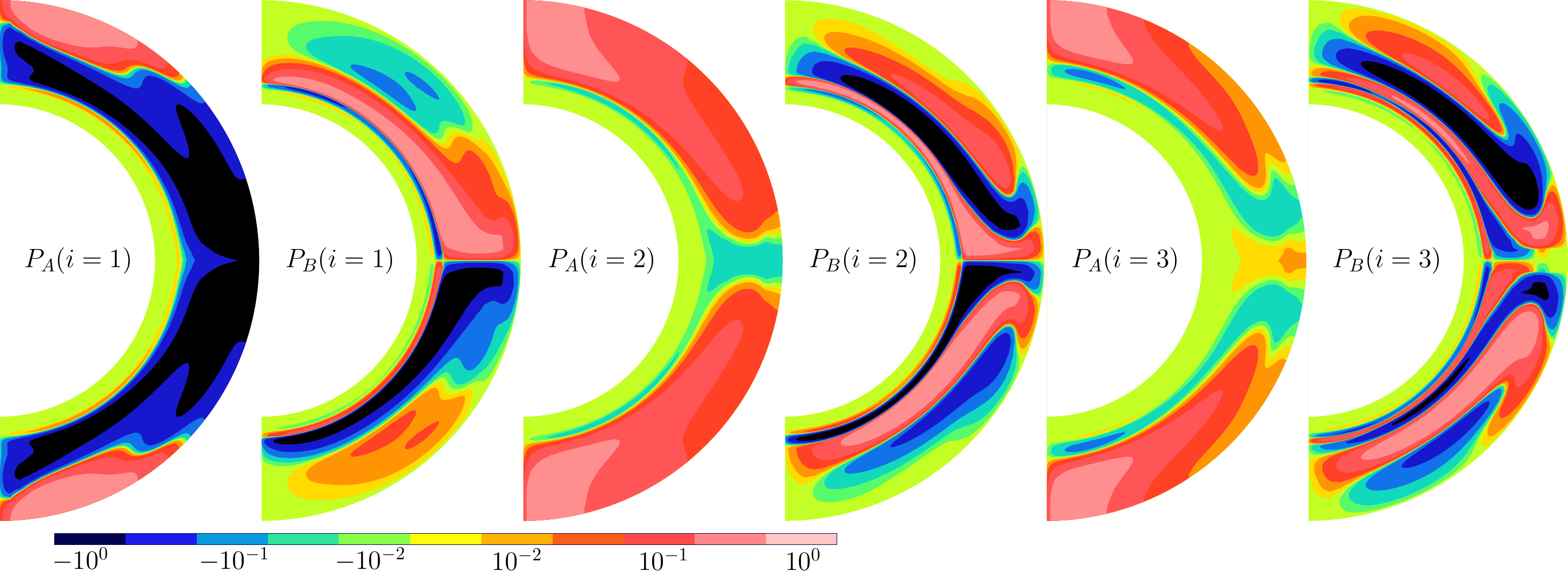}
\caption{First 3 eigenfunctions (with leading eigenvalues) of the covariance matrix of a dynamo model based on a unicellular flow,
expressed on the meridional plane. $P_A$ (resp. $P_B$) is the poloidal (resp. toroidal) component of the eigenvector. 
The higher modes with lower eigenvalues display more structures on the meridional plane. }
\label{fig:eigenfcns}
\end{figure}

\begin{figure}[!ht]
\includegraphics[draft=\draftgraphicx,width=0.7\columnwidth]{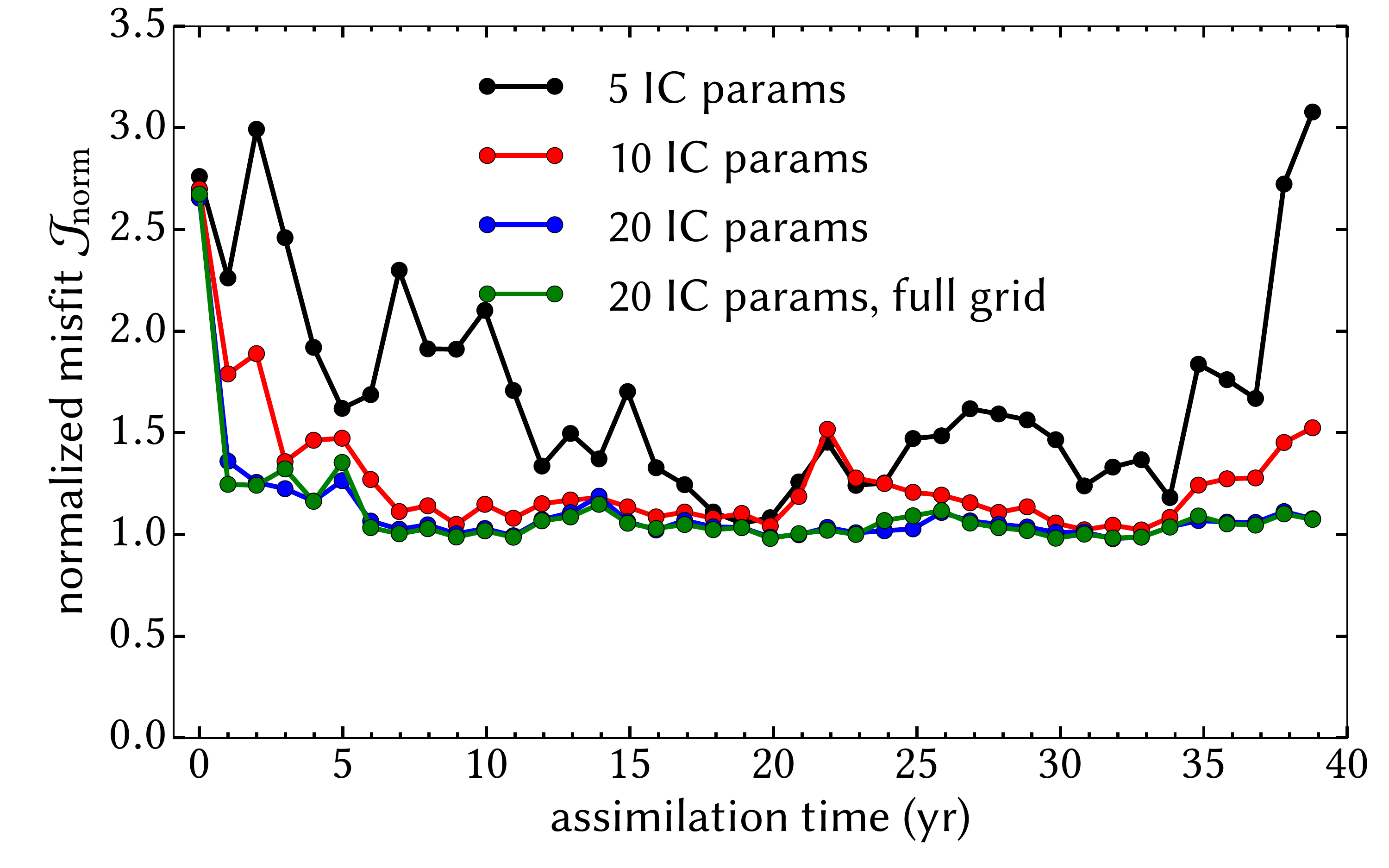}
\caption{
Normalized misfit versus time over the course of the assimilation
 for different parameterizations of the magnetic component
of the control vector. Black (resp. red, resp. blue): 5 (resp. 10, resp. 20) eigenmodes are
retained after the diagonalization of the covariance matrix constructed from 
 the knowledge of the magnetic field at every other grid point. Green: 20 eigenmodes
 are retained after the diagonalization of the magnetic covariance matrix constructed
 from the knowledge of the magnetic field at every grid point. } 
\label{fig:NmisfitplotIC}
\end{figure}

To reduce the size of computation, we only include every other 
grid point in $r$ and $\theta$ in the construction of the field vector and covariance matrix, 
and interpolate in the coordinate space to approximate the magnetic field. 
This reduces the size of $\mathbf{P}$ by a factor of 16.
As shown in Fig. \ref{fig:NmisfitplotIC}, 
this has little impact on the scheme, whose accuracy is mostly controlled by the level of noise
 impacting the data. 
In addition to the spectrum and eigenbasis of the covariance matrix we showed in Fig. \ref{fig:ecovar} of Sec. \ref{subsec:assimsetup}, 
 we also show the physical structure of the first 3 eigenfunctions in Fig. \ref{fig:eigenfcns}:  
Higher modes with lower eigenvalues display more complex structures in the meridional plane. 
\par
The spectrum of the covariance matrix (Fig. \ref{fig:ecovar} (a)), 
and the error in the approximation of a dynamo field by a truncated basis of eigenvectors 
(Fig. \ref{fig:ecovar} (b)) drop rapidly with the level of truncation. 
The error in Fig. \ref{fig:ecovar} (b) flattens to a few  percents,  a consequence
of the every other point approximation discussed above; again, this  
approximation does 
 not impact the overall accuracy of the scheme, which is controlled by the observational noise. 

The forward model is initialized with such a representation, 
and the corresponding adjoint operator is developed similarly. 
(Recall that the derivative of the objective function with respect to the initial field 
is the corresponding adjoint field at the beginning of the assimilation window \hcpi{Paper I}.) 
The covariance matrix in the $n^{th}$ step 
is evaluated from the dynamo model forecast in the $(n-1)^{th}$ assimilation window. 
For $n=1$, the dynamo model 
is a simple one based on unicellular flow. 
Updating of the covariance matrix after each year can ensure 
we can capture the change in the dynamics and statistics of the dynamo action, 
and in consequence the initial conditions can be reasonably approximated.
\par

\begin{figure}[!ht]
\includegraphics[draft=\draftgraphicx,width=.6\columnwidth]{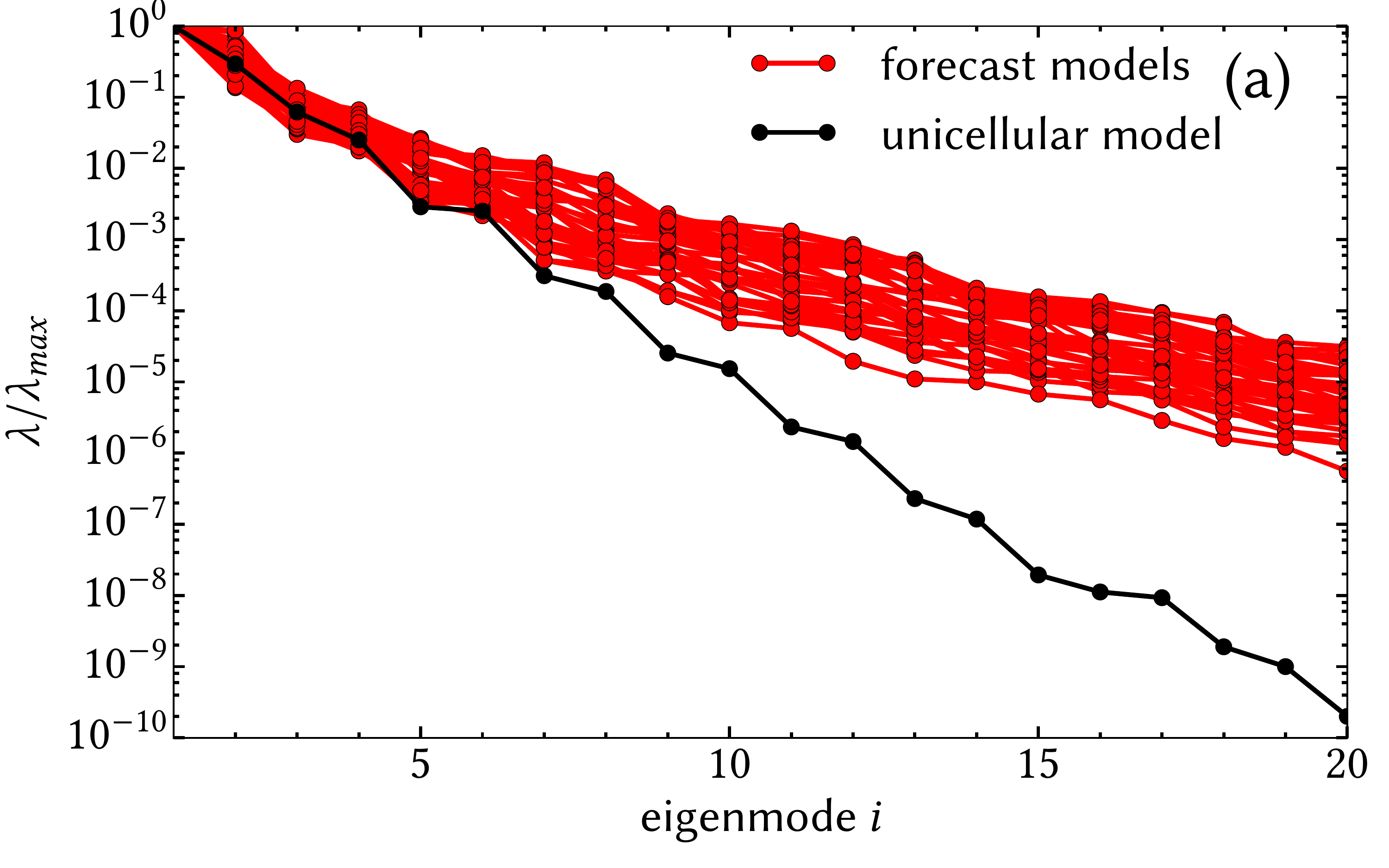}\\
\includegraphics[draft=\draftgraphicx,width=.6\columnwidth]{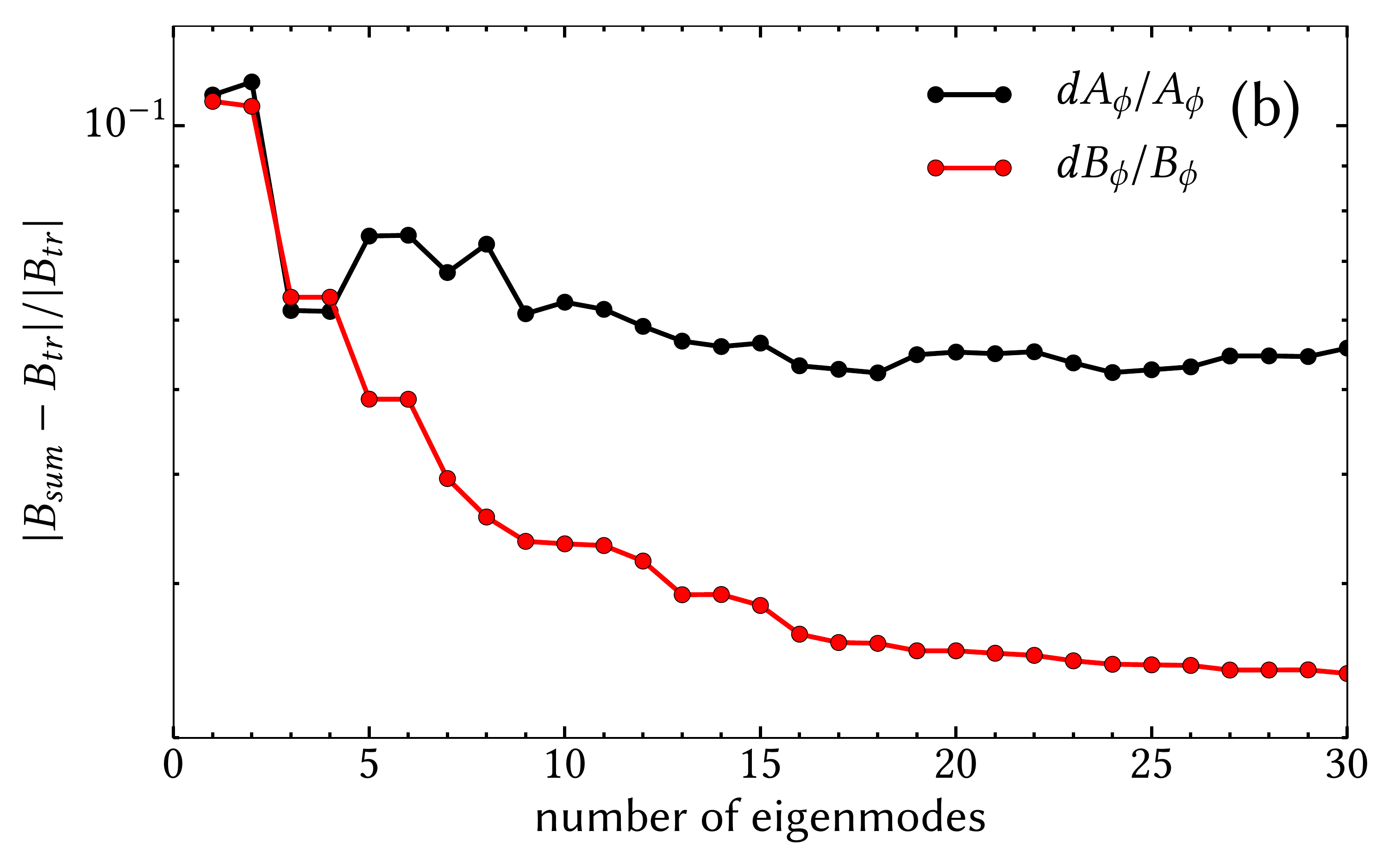}
\caption{(a) Red: Eigenvalue spectra of the covariance matrices of the hind-cast dynamo models 
for the estimated flow in each of the 40 years assimilation. The corresponding spectrum of 
the dynamo model for a unicellular steady flow is shown again in a black curve for comparison.
(b) Error in the approximation of the hind-cast magnetic field at the end of the assimilation at the 40th year, 
as a function of the size of a truncated eigenbasis. Black: error in the poloidal field. 
Red: error in the toroidal field.
}
\label{fig:finalerrorfore}
\end{figure}

In Sec. \ref{subsec:assimsetup} we truncate the expansion of the initial condition to $m=20$ leading eigenvectors, 
as the spectrum of $\mathbf{P}$ and the of error in expanding 
a simple dynamo field drop rapidly when the mode number increases (Fig. \ref{fig:ecovar}). 
To justify this approximation, 
we perform the assimilation experiment of our reference case, at various $m$,
and using a simple unicellular flow as the prior for the first year of assimilation.
We show the misfit in Fig. \ref{fig:NmisfitplotIC}.
We can see that at $m=20$, we have an optimal misfit of $\sim 1$, and for more aggressive truncation
of the eigenbasis representation, there will be underfitting.
The size of the truncated basis required is related to the spectrum of the covariance matrix. 
In Fig. \ref{fig:finalerrorfore} (a), we can see that the covariance matrices for the models forecast 
during the 40 years of assimilation give broader spectra compared with a unicellular prior.
This means higher eigenmodes are more important for more complicated magnetic configuration
as the assimilation procedure proceeds. 
To illustrate that, we show the error in expanding the forecast magnetic
field at the end of the assimilation of 40 years with the eigenbasis of the final forecast model 
in Fig. \ref{fig:finalerrorfore} (b). Compared with Fig. \ref{fig:ecovar} (b), the error converges at 
higher $m$, but still soundly contained in our chosen size $m=20$. 
Therefore, we justify the truncation of the eigenbasis in representation 
of the initial condition at $m=20$ in our tests. 

To summarize, 
the procedures of the data assimilation for our course of 40 years analysis of synthetic observations 
are listed as follow:
\begin{enumerate}
\item For $n=1$, calculate the covariance matrix $\mathbf{P}_1(t_{s,1})$ of the initial guess of the initial conditions.
(Here $t_{s,n}$ and $t_{e,n}$ are respectively the starting time and the ending time of the assimilation window at the $n^{th}$ step 
and we have $t_{s,n}=t_{e,n-1}$.)
Usually the guess is the dynamo model based on unicellular flow with magnetic cycle of 22-years. Diagonalize the covariance matrix 
and project the guess of initial magnetic field on the eigenbasis to obtain $\mathrm{x}^g_{1,IC}$, the superscript $g$ stands for guess. 
Combined with the guess of the meridional flow $\mathrm{x}^g_{1,MC}$, we have $\mathbf{x}^g_1$ for assimilation 
of the observations of the first year.
\item Based on the synthetic observations of the first year, with an appropriate guess $\mathbf{x}^g_1$, 
the data assimilation procedure gives a forecast of magnetic field, and an analyzed meridional flow $\mathbf{x}^f_1$, 
the superscript $f$ stands for forecast.
\item For $n > 1$, construct the covariance matrix $\mathbf{P}_n(t_{e,n-1})$ of the dynamo model based on the analyzed flow at the $n-1$ 
assimilation $\mathrm{x}^f_{n-1,MC}$.
Evaluate the eigenbasis of $\mathbf{P_n}(t_{e,n-1})$, and project the analyzed magnetic field from the assimilation
window $n-1$ at $t_{e,n-1}$ and obtain $\mathrm{x}^g_{n,IC}$. 
The initial guess of the flow in step $n$ will be the analyzed flow in
step $n-1$, i.e., $\mathrm{x}^g_{n,MC}=\mathrm{x}^f_{n-1,MC}$. 
So we obtain $\mathbf{x}^g_n$. 
\item Based on the synthetic observations at the $n^{th}$ year, with the guess $\mathbf{x}^g_n$, the data assimilation procedure 
gives the analysis $\mathbf{x}^f_n$. 
The analyzed magnetic field and the estimated flow will give the initial guess $\mathbf{x}^g_{n+1}$ 
and so on and so forth until $t_{e,40}$ is reached.
\end{enumerate}


\section{Brief analysis of temporal variability of meridional flow}
\label{sec:anassn}
In this section we present an analysis of the surface flow of the Sun 
which shows the temporal variability, 
using data from \cite{Ulrich2010}.


\begin{figure}[!ht]
\includegraphics[draft=\draftgraphicx,width=0.7\columnwidth]{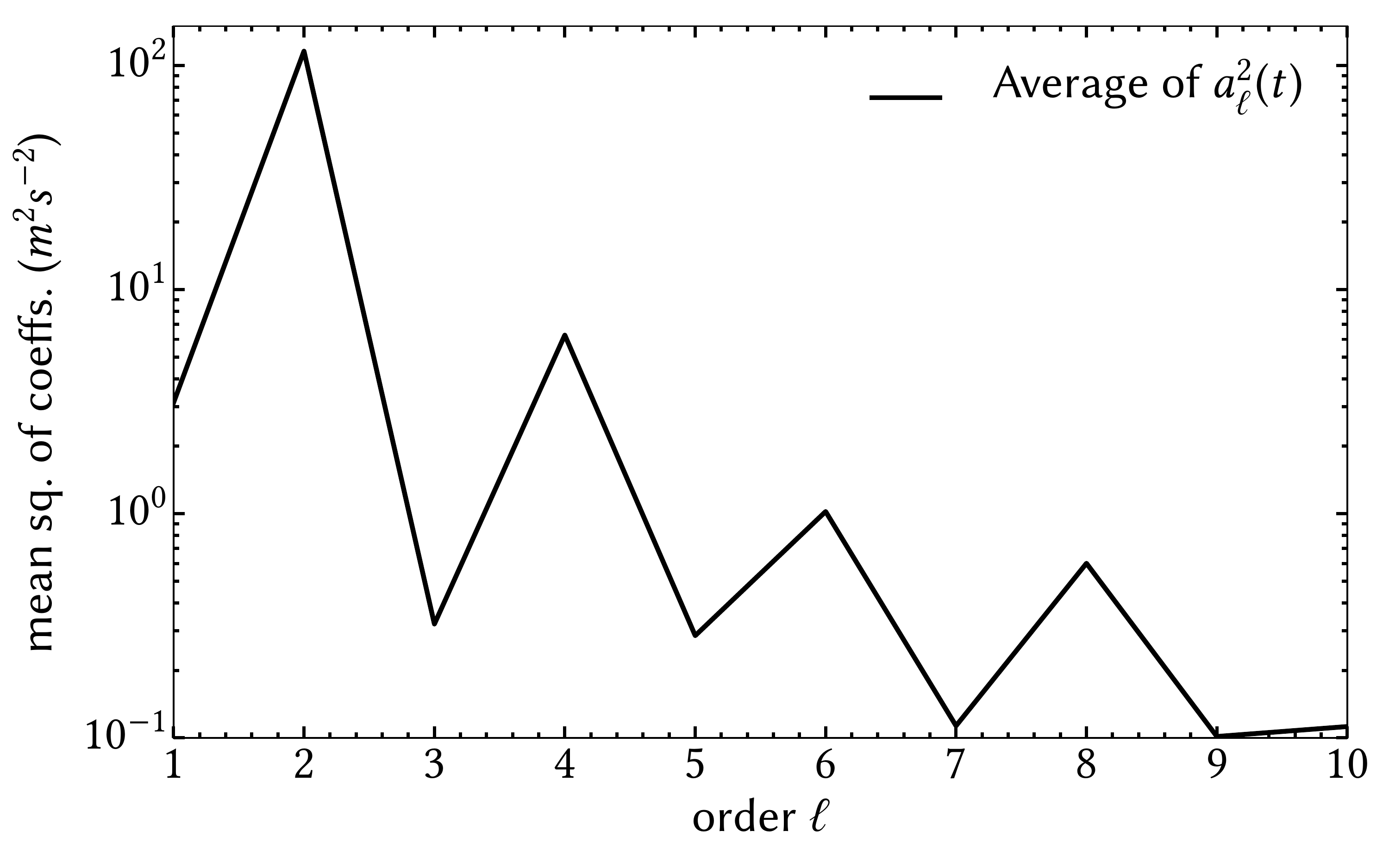}
\caption{The mean square (in time) of the expansion coefficients of the surface flow 
as a function of the degree $\ell$ of associate Legendre polynomials of order $1$, $P^1_\ell$. 
}
\label{fig:p1jave}
\end{figure}

\begin{figure}[!ht]
\includegraphics[draft=\draftgraphicx,width=0.7\columnwidth]{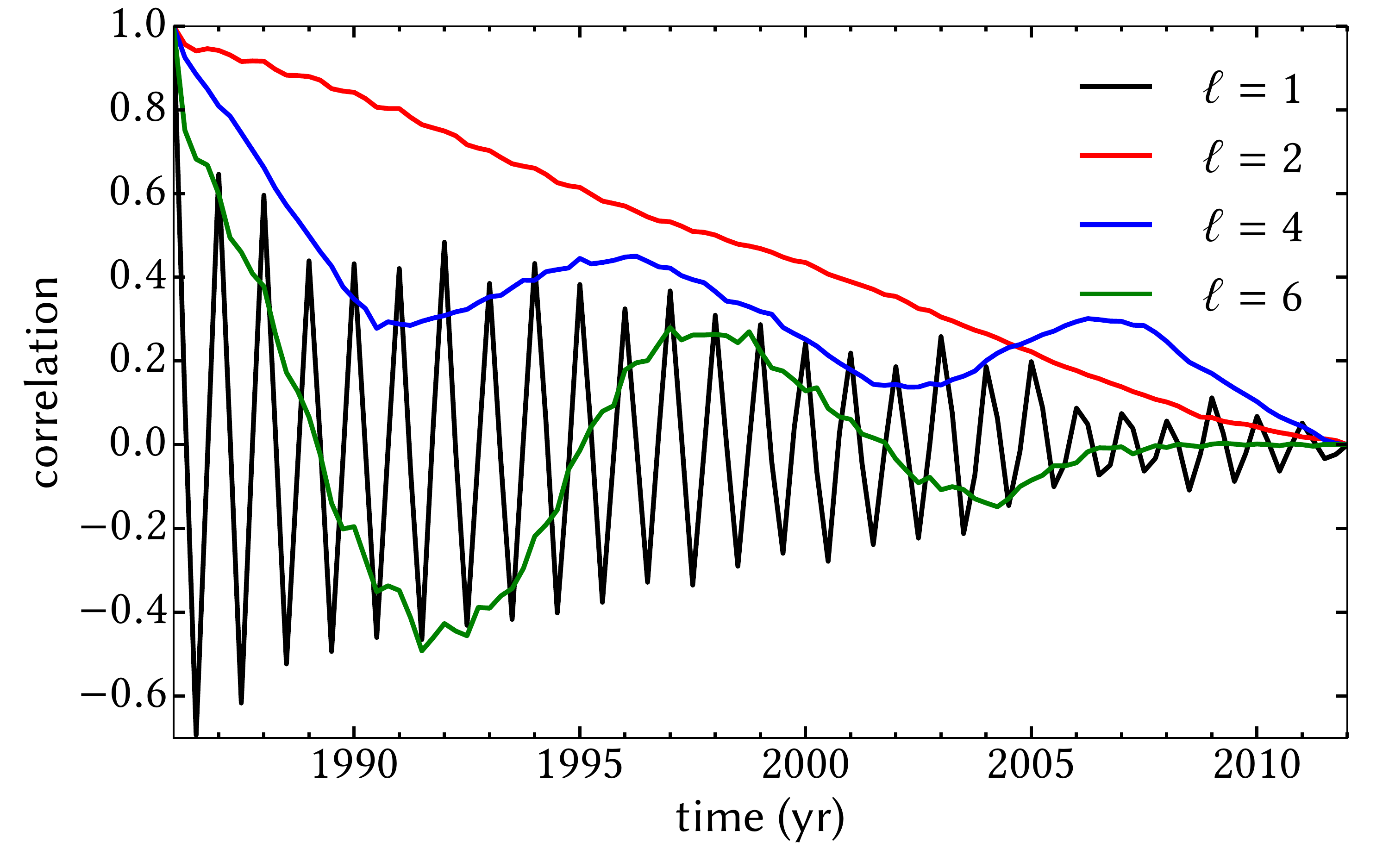}
\caption{The autocorrelation functions of the time series of the expansion coefficients 
of the surface flow (on $P^1_\ell$).
Only the 4 coefficients with highest mean square in time average are shown 
($\ell=$ 1 (black), 2 (red), 4 (blue), 6 (green)). 
}
\label{fig:autoj14}
\end{figure}
In Sec. \ref{subsec:modeleq}, we mentioned that the correlation time of the spectrum of the surface 
meridional flow is of order 1 year. 
The observed flow on the solar surface can be found, for example, in \cite{Ulrich2005, Ulrich2010}.
The flow is dominantly poleward at the surface.
We project the flow on the associated Legendre polynomials of order 1 ($P^1_\ell$), 
and plot the mean square (in time) of the spectrum in Fig.~\ref{fig:p1jave}. 
The modes which are odd about the equator, 
i.e. with even $\ell$, are dominant over their even parity counterparts, 
and the spectrum in general decreases with increasing $\ell$.
\par
To study the temporal variability of the flow, we evaluate the auto-correlation of the expansion coefficients 
on $P^1_\ell$s, for $\ell = 1,2,4,6$ and show it in Fig.~\ref{fig:autoj14}. The first 3 equatorially odd modes 
display correlation times of at least 5 years, and the first equatorially even counterpart $\ell=1$, is of 
correlation times $\sim 1$ year. We have thus decided to use a modulation for the flow of 3 years 
as illustrated in Fig. \ref{fig:timevaryflow} and \ref{fig:uthetacontour}, which results in 
time dependent modulation of the flow in good agreement with observations.

\acknowledgments
\begin{center}
  {ACKNOWLEDGMENTS}
\end{center}
We acknowledge financial support of the UnivEarthS Labex program at Sorbonne-Paris-Cit\'{e} 
(ANR-10-LABX-0023 and ANR-11-IDEX-0005-02) through project SolarGeoMag. 
We also acknowledge the support from  
the ERC PoC SolarPredict project, CNES Solar Orbiter and INSU/PNST grants, and Idex SPC through the DAMSE project.
We are grateful to Roger Ulrich for giving us digital access to 
his surface meridional circulation measurements, 
we also thank SIDC for access to their sunspot series observations. 
Wilcox Solar Observatory data used in this study was obtained via 
the web site http://wso.stanford.edu at $2017:09:29$ $03:15:25$ PDT courtesy of J.T. Hoeksema. 
ASB thanks M. DeRosa  and A. Title for useful discussions.
Numerical computations are performed on the S-CAPAD platform, IPGP, France and
on DIM-ACAV funded Anais server at CEA/IRFU. 
IPGP contribution nnnn.  \\


\bibliography{mctvf}
\bibliographystyle{./apj}

\end{document}